\documentclass[aps,prd,groupedaddress,nofootinbib,english,nobalancelastpage,floatfix,superscriptaddress]{revtex4-1}
\pdfoutput=1
\usepackage{flushend}
\DeclareMathAlphabet{\scr}{U}{rsfs}{m}{n}
\usepackage{graphicx}
\usepackage{amsmath,bm}
\usepackage{multirow}
\usepackage{rotating}
\usepackage{float}
\usepackage[normalem]{ulem}

\usepackage[usenames,dvipsnames]{xcolor}
\definecolor{naviBlue}{RGB}{0,0,128}
\usepackage[colorlinks  = true,
            linkcolor   = naviBlue,
            urlcolor    = naviBlue,
            citecolor   = naviBlue,
            anchorcolor = naviBlue]{hyperref}

\newcommand{\D}{\mathrm{d}}
\newcommand{\half}{\frac{1}{2}}

\def\bea{\begin{eqnarray}}
\def\eea{\end{eqnarray}}
\def\be{\begin{equation}}
\def\ee{\end{equation}}

\newcommand{\io}[2]{#1_{#2}}
\newcommand{\is}[3]{#1_{#2,#3}}
\newcommand{\ia}[4]{#1^{#2}_{#3,#4}}

\newcommand{\RB}[1]{\Big( #1 \Big) }
\newcommand{\SQB}[1]{\bigg[ #1 \bigg] }

\newcommand{\rb}[1]{\left( #1 \right) }

\def \aap  {A\&A}

\def \apj  {ApJ}
\def \apjs  {ApJS}

\def \mnras {MNRAS}

\def \pasp {PASP}

\begin{document}
\title{The self-confinement of electrons and positrons from dark matter}

\author{Marco Regis} % [0000-0003-0399-0284]
\email{marco.regis@unito.it}
\affiliation{Dipartimento di Fisica, Universit\`a di Torino, Via P. Giuria 1, 10125 Torino, Italy}
\affiliation{Istituto Nazionale di Fisica Nucleare, Sezione di Torino, Via P. Giuria 1, 10125 Torino, Italy}

\author{Michael Korsmeier} % [0000-0003-3478-888X]
\email{michael.korsmeier@fysik.su.se}
\affiliation{The Oskar Klein Centre, Department of Physics, Stockholm University, AlbaNova, SE-10691 Stockholm, Sweden}

\author{Gianni Bernardi}
%\email{gianni.bernardi@inaf.it}
\affiliation{INAF - Istituto di Radioastronomia, via Gobetti 101, 40129 Bologna, Italy}
\affiliation{Department of Physics and Electronics, Rhodes University, PO Box 94, Grahamstown, 6140, South Africa}
\affiliation{South African Radio Astronomy Observatory, Black River Park, 2 Fir Street, Observatory, Cape Town, 7925, South Africa}

\author{Giada Pignataro}
%\email{giada.pignataro@inaf.it}
\affiliation{Dipartimento di Fisica e Astronomia, Universit\`a degli Studi di Bologna, via P. Gobetti 93/2, 40129 Bologna, Italy}
\affiliation{INAF - Istituto di Radioastronomia, via Gobetti 101, 40129 Bologna, Italy}

\author{Javier Reynoso-Cordova}
%\email{javierisreal.reynosocordova@unito.it}
\affiliation{Dipartimento di Fisica, Universit\`a di Torino, Via P. Giuria 1, 10125 Torino, Italy}
\affiliation{Istituto Nazionale di Fisica Nucleare, Sezione di Torino, Via P. Giuria 1, 10125 Torino, Italy}

\author{Piero Ullio}
%\email{ullio@sissa.it}
\affiliation{Scuola Internazionale Superiore di Studi Avanzati (SISSA), via Bonomea 265, 34136 Trieste, Italy}
\affiliation{INFN, Sezione di Trieste, via Valerio 2, 34127 Trieste, Italy}
\affiliation{Institute for Fundamental Physics of the Universe (IFPU), via Beirut 2, 34151 Trieste, Italy}

\begin{abstract}
Radiative emissions from electrons and positrons generated by dark matter (DM) annihilation or decay are one of the most investigated signals in indirect searches of WIMPs.
Ideal targets must have large ratio of DM to baryonic matter. However, such ``dark'' systems have a poorly known level of magnetic turbulence, which determines the residence time of the electrons and positrons and therefore also the strength of the expected signal. This typically leads to significant uncertainties in the derived DM bounds.
In a novel approach, we compute the self-confinement of the DM-induced electrons and positrons. Indeed, they themselves generate irregularities in the magnetic field, thus setting a lower limit on the presence of the magnetic turbulence.
We specifically apply this approach to dwarf spheroidal galaxies.
Finally, by comparing the expected synchrotron emission with radio data from the direction of the Draco galaxy collected at the Giant Metre Radio Telescope, we show that the proposed approach can be used to set robust and competitive bounds on WIMP DM.
\end{abstract}

\maketitle

\section{Introduction}
\label{sec:intro}

Dark matter (DM) is a fundamental ingredient in the formation of structures in our Universe.
Yet, its fundamental nature remains elusive. 
Weakly interacting massive particles (WIMPs) are one of the most investigated classes of DM candidates in the literature~\cite{Bertone:2010zza}. The weak interaction implies that WIMPs in DM halos annihilate in pairs or decay into detectable species. Thus, information on the WIMP physical properties can be obtained through indirect searches, namely by studying the astrophysical signals associated to the annihilation/decay. Among the various channels, a sizable production of electrons and positrons is a general feature of WIMP models~\cite{Regis:2008ij}. They in turn generate synchrotron emission in the magnetized atmosphere of astrophysical structures, which could be observed as a diffuse radio emission centered on the DM halo distribution~\cite{Colafrancesco:2005ji}.

The search for faint, diffuse radio emission in targeted galaxies is expected to undergo a golden era in the forthcoming years. New radio facilities, like LOFAR, the GMRT, ASKAP and MeerKAT, have started operation. In the next decade, the SKAO will be the radio telescope providing unmatched sensitivity, angular resolution and frequency coverage, to search in a deeper and deeper way for diffuse radio emissions~\cite{SKAMagnetismScienceWorkingGroup:2020xim}.

It looks appealing to search for WIMP signatures with radio telescopes~\cite{Regis:2017oet,Regis:2021glv}. An important obstacle immediately appears, and it is the presence of radio diffuse emission from ``ordinary" astrophysical processes. How to disentangle it from the possible DM contribution? The conceptually simplest way to address the issue is to focus on dark, quiescent systems, where the star formation rate is suppressed, and thus cosmic-ray (CR) accelerators that would lead to ultra-relativistic electrons and to synchrotron radiation are absent. 

According to numerical simulations, the halos of galaxies are populated by smaller systems, with many of such subhalos being inefficient in forming stars. A population of satellites that is observationally detected and that lacks recent star formation is given by dwarf spheroidal (dSph) galaxies~\cite{McConnachie:2012vd}.
A key question about a possible synchrotron emission from dark systems is about the presence of a significant magnetic field. While it is plausible to assume a coherent component at the level of $\mu$G~\cite{Regis:2014koa}, either generated during the history of the object or coming from the large scale magnetic lines of the host galaxy, the presence of a turbulent component, in absence of star formation, is more uncertain. The turbulence is fundamental in setting the confinement scale of the electrons and positrons. This is a crucial point for the WIMP signal since the objects we just mentioned are typically quite small, and electrons might travel well outside the object region before emitting a significant synchrotron radiation.

Such uncertainty on the magnetic properties has led to an uncertainty of several orders of magnitude in the predicted radio signal in dSphs~\cite{Regis:2014tga}, and thus in the bounds on the WIMP annihilation rate~\cite{dwNatarajan:2013,dwSpekkens:2013,dwNatarajan:2015,Regis:2014tga,Regis:2017oet,dwKar:2019,dwCook:2020,dwVollmann:2020a,dwBasu:2021,dwGajovic:2023bsu}. Its assessment is thus mandatory to determine whether or not dSphs are promising targets for WIMP radio searches.

In this work, we propose a novel approach, which considers the fact that the electrons and positrons injected by WIMPs generate themselves magnetic irregularities and so they induce a certain level of turbulence. We quantitatively compute such effect and derive under which conditions this mechanism is relevant to effectively confine $e^+-e^-$ inside ``dark" systems.

The paper is organized as follows. Section~\ref{sec:eqs} describes the diffusion and turbulence equations and their solution (with numerical details in Appendix~\ref{sec:num} and an approximate analytical solution in Appendix~\ref{sec:est}). The formalism is then applied to the case of dSph galaxies in Section~\ref{sec:dSph}, where we also describe the impact of the magnetic field modeling on the estimate of WIMP-induced synchrotron radiation. In Section~\ref{sec:radio}, we focus on a specific case, the Draco dSph, and we derive bounds on WIMP DM  by comparing the expected emission with observations from the GMRT telescope. Details on the DM profile in Draco and on the radio observations are reported in Appendix~\ref{sec:jeans} and \ref{sec:obs}. We summarize our findings in Section~\ref{sec:conc}.

\section{Evolution of the electron/positron density and magnetic turbulence}
\label{sec:eqs}
The propagation of CRs in turbulent magnetic fields is an inherently difficult problem~\cite{Berezinsky:1990qxi}. The charged CRs are deflected in the magnetic fields. In a constant magnetic field, CRs spiral around the magnetic field lines. However, as can be shown in linear pertubation theory, the presence of a (small) turbulent magnetic field on top of the regular field, randomly changes the angle between the motion of the CR and the regular magnetic field, effectively leading to the diffusion of the CR along the magnetic field lines. At the same time, the motion of the charged CRs induces magnetic fields and stimulates turbulence~\cite{Cesarsky:1980pm}. The equations describing the coupling of magnetic fields to the CR plasma are studied in magnetohydrodynamics. 
While there are numerical tools (e.g., particle-in-cell simulations) to solve these equations, the currently available computing power is not sufficient for the scale of (dwarf) galaxies, given the several orders of magnitude of separation between the galaxy scale and the CR gyro-radius. Therefore, we rely on a more phenomenological-driven method, where the transport of electrons and positrons is described by diffusion, advection, and energy losses. Then we assume spherical symmetry. This is a simplifying but fair approximation, since representative of a model in which a strong gradient in the CR intensity may manifest in connection to a gradient in the source function and one in the turbulence distribution. On the other hand, the two gradients may not be aligned (e.g., a spherical source versus turbulence flowing along lines of a regular magnetic field being large scale dipole or toroidal).

With above assumptions, the transport equation is given by \cite{Blandford:1987pw}:
\begin{equation}
\label{eq:neq}
    \frac{\partial n_e}{\partial t}
    = \frac{1}{r^2}\frac{\partial }{\partial r}\SQB{r^2 D\frac{\partial n_e}{\partial r}-r^2 v_A n_e}
    + \frac{2 v_A}{r}
      \frac{\partial}{\partial E}\SQB{ \frac{p}{3} \beta\,c\, n_e   }
    - \frac{\partial }{\partial E} \SQB{ \dot E n_e }
    + q_{\rm CR}
     \, .
\end{equation}
Here, $n_e$ is the electron (or positron) density in the energy interval $[E, E+dE]$, $D$ is the spatial diffusion coefficient which may depend on the radial distance $r$ and energy $E$, $v_A$ is the Alfv\'en velocity that transports the electrons radially away from the center of the system, $c$ is the speed of light, $q_{\rm CR}$ is the source term, and $\dot E$ describes the energy losses due to synchrotron emission and inverse Compton scattering.%
\footnote{We note that there are two different ways to write the advection term in literature. The divergence term can be partly reshuffled between electron current (first term) and the adiabatic energy-losses (second term).}

\begin{figure}[t]
\centering
\vspace{-4cm}
\includegraphics[width=0.6\textwidth]{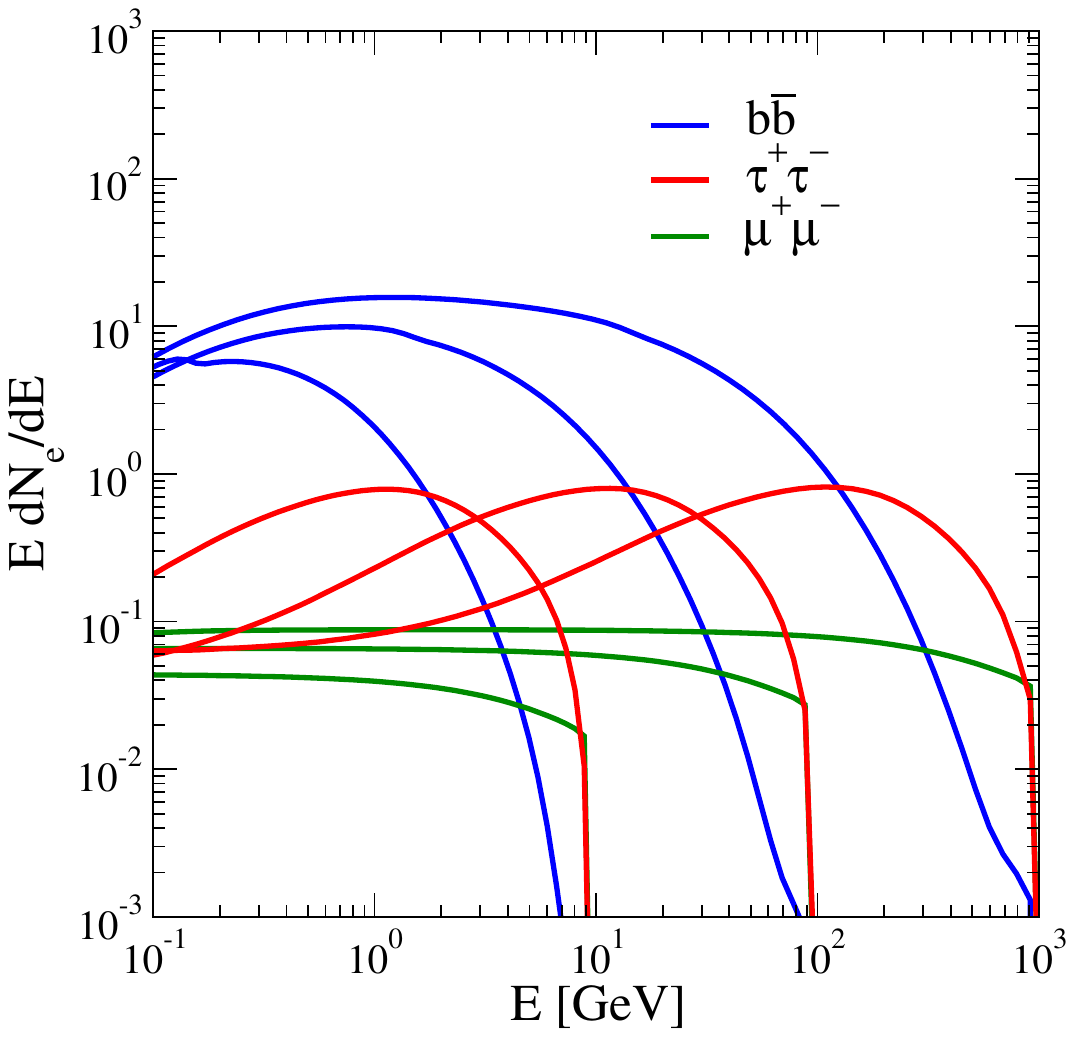}
    \caption{ Injection spectra of $e^+/e^-$ for different annihilation channels,  $b\bar b$ (blue), $\tau^+\tau^-$ (red) and $\mu^+\mu^-$ (green), and for three different DM masses, 10 GeV, 100 GeV and 1 TeV (see cutoff). }
\label{fig:dNdE}
 \end{figure}

In this work we consider electrons and positrons to be injected by the pair annihilation of DM particles. We assume that there is no other source coming from astrophysical processes. Thus, the source term is given by:
\begin{equation}
\label{eq:qc}
    q_{\rm CR}(r,E) = \langle \sigma v \rangle_f 
                        \frac{\rho_{\rm DM}^2(r)}{2\,m_{\rm DM}^2} 
                        \frac{dN_e^f}{dE}\; ,
\end{equation}
where $\rho_{\rm DM}$ is the DM energy density, $m_{\rm DM}$ is the DM mass, $\langle \sigma v \rangle_f$ is the velocity-averaged annihilation cross section into the final state $f$, and ${dN_e^f}/{dE}$ is the energy spectrum of the electrons and positrons produced from the annihilation, which depends on the annihilation channel $f$. In Fig.~\ref{fig:dNdE}, we show a few examples of ${dN_e^f}/{dE}$ used in this work. We model $\rho_{\rm DM}$ with a Navarro-Frenk-White (NFW) DM profile~\cite{Navarro:1995iw}: $\rho_{\rm NFW}(r)=\rho_s\,r_s/r/(1+r/r_s)^2$, with two free parameters, $\rho_s$ and $r_s$. In Appendix~\ref{sec:jeans} we discuss also a different choice, considering a cored profile.

The advection term depends on the Alfv\'en speed that, for a given magnetic field, can be estimated from the plasma density $n_p$ through $v_A\simeq 63\,{\rm km/s}\,\sqrt{10^{-3}{\rm cm^{-3}}/n_p}\,B_0/\mu{\rm G}$.

\begin{figure}[b!]
\includegraphics[width=0.49\textwidth]{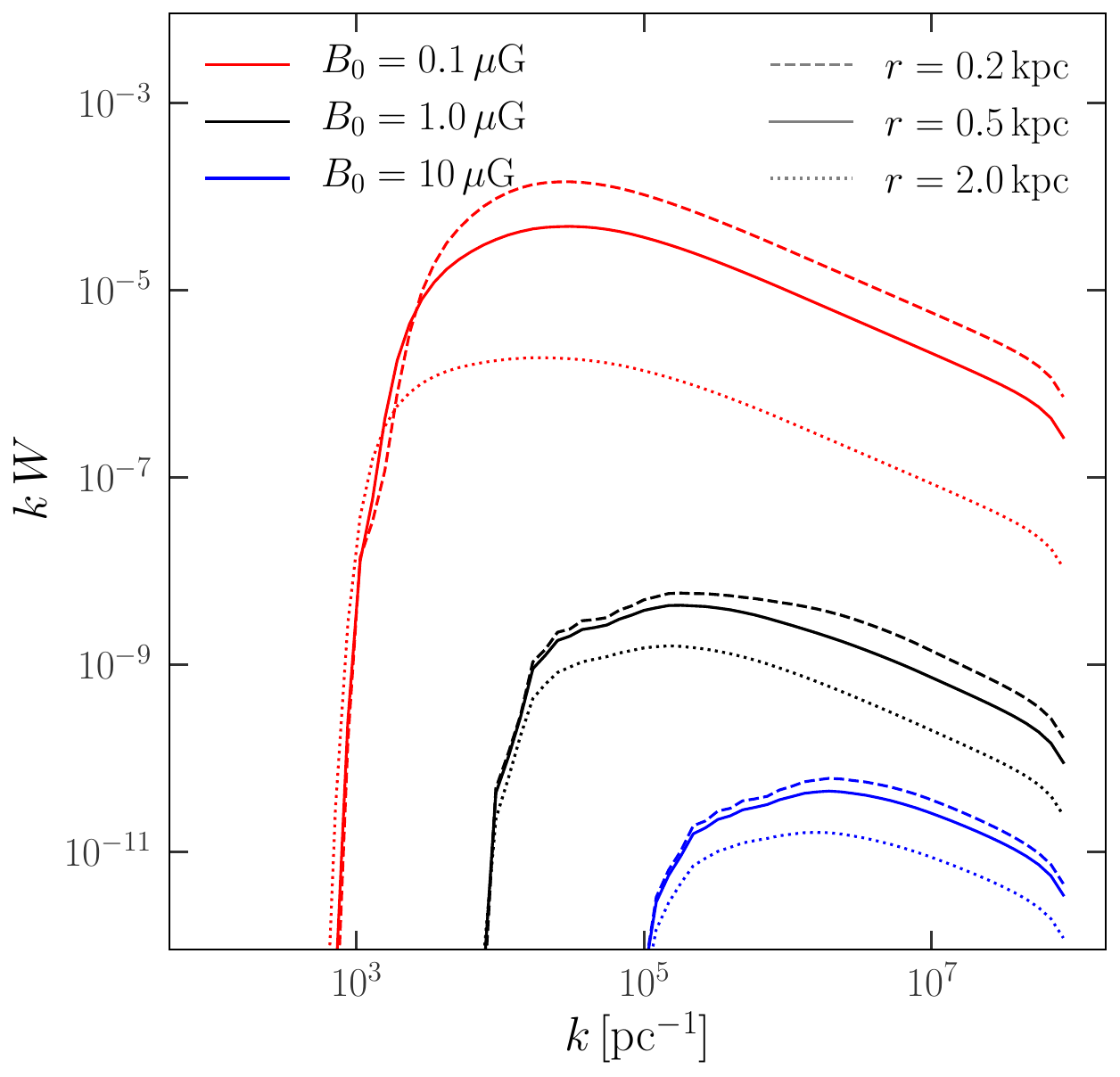}%
\includegraphics[width=0.48\textwidth]{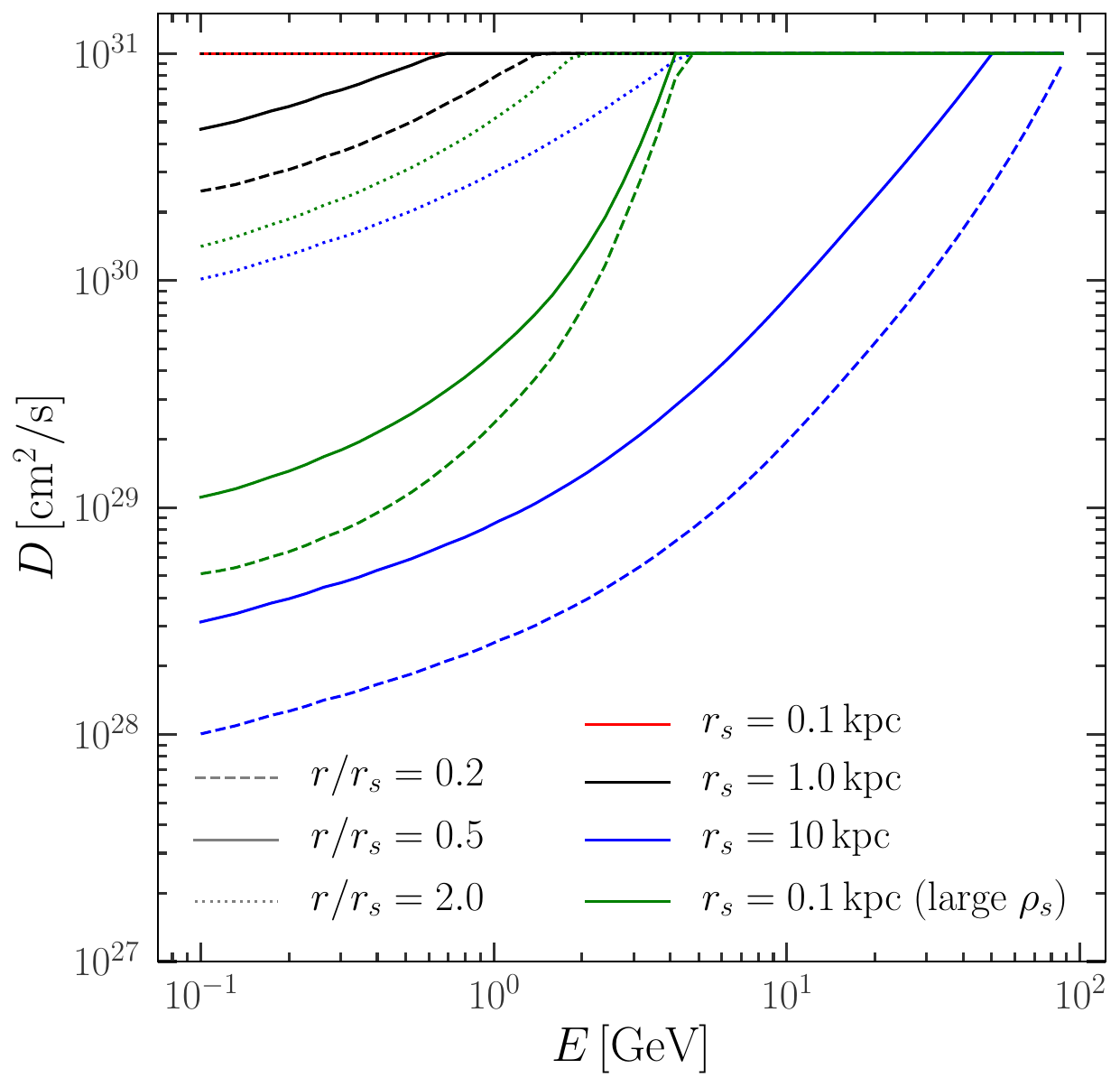}
    \caption{Left: Power spectrum of the magnetic turbulence as a function of the wave number, for a reference DM scenario (detailed in the text) and at three different distances from the center of the system, 0.2 kpc (dashed), 0.5 kpc (solid), and 2.0 kpc (dotted). We show the effect of varying the strength of the regular magnetic field, considering $B_0=0.1\,\mu$G (red), $B_0=1.0\,\mu$G (black), and $B_0=10\,\mu$G (blue). We remind that $E \simeq (10^6{\rm pc^{-1}}/k)\,(B_0/\mu{\rm G})$ GeV. Right: Spatial diffusion coefficient as a function of energy for the same particle DM scenario and radial distances of the left panel. We investigate the dependence on the size and normalization of the DM profile by considering $\rho_s=4\times 10^7\,M_\odot/{\rm kpc}^3$ with $r_s=0.1,\,1,\,10$ kpc (red, black, blue) and $\rho_s=4\times 10^8\,M_\odot/{\rm kpc}^3$ with $r_s=0.1$ kpc (green).}
\label{fig:W_example}
 \end{figure}

The diffusion of CRs is mediated by scattering processes on the turbulent magnetic field and we denote the power spectrum of the magnetic turbulence with $W = W(k, r, t)$. From linear perturbation theory, the relation between the diffusion coefficient and the turbulence spectrum is given by~\cite{Berezinsky:1990qxi}: 
\begin{equation}
\label{eq:dc}
    D(r, p, t)=\frac{D_{\rm B}(p) 4/\pi}{kW(r,k,t)} \; ,
\end{equation}
where
$D_{\rm B}(p)=r_ {\rm L}(p)c\beta/3$ is the Bohm diffusion coefficient. 
For resonant interaction between Alfv\'en waves and CR scattering, the wave number $k$ and the momentum $p$ of the CR are related by via the Larmor radius, $r_ {\rm L}(p_{\rm res}) = 1/k_{\rm res}\simeq 3.3\times 10^7\,(E/{\rm GeV)}\,(\mu{\rm G}/B_0)$ km.

Turbulence in a galaxy with sizeable star formation rate, such as the Milky Way, is usually assumed to be injected in the system as an astrophysical feedback, such as from supernova explosions~\cite{2013A&ARv..21...70B}; on the other hand, recent reanalyses - mostly in connection to spectral features in local CR fluxes - have shown that streaming instabilities of CR themselves may be an additional and relevant source of turbulence in the Galaxy~\cite{Blasi:2012yr,Evoli:2018nmb,Jacobs:2021qvh}. 
Self-generation of magnetic turbulence is also studied as a solution to the observed inhibited diffusion in TeV halos around pulsar wind nebulae \cite{Evoli:2018aza,Mukhopadhyay:2021dyh}. Here the turbulence is created by the injection of the electrons and positrons from the pulsar with a burst-like source term.
%instead of from DM annihilation. The equations describing the system are very similar, however, there are systematic differences between DM halos and pulsars, as for example a continuous source term versus a burst-like injection of the leptons.

Considering ``dark” systems, like dwarf spheroidal galaxies or isolated DM clumps, in which the star formation rate is highly suppressed, we discuss here the possibility that this second effect alone can become sufficiently efficient to give rise to a diffusive halo. 
In this scenarios, we assume that turbulence is self-generated only by the electrons and positrons from DM annihilation. 
This is clearly a lower bound for the turbulence (since some rare astrophysical events might still contribute) and thus a conservative estimate. In particular, if giant halos around galaxies exist, as possibly suggested by $\gamma$-ray data~\cite{Gabici:2021rhl}, some non-negligible CR and plasma densities might be present at the dSph location, leading to non-negligible generation of turbulence and damping. Also, depending on the annihilation final states, there might also be a significant protons/antiprotons production from DM, which would enhance the turbulence. Including these additional contributions is beyond the scope of this work, which focuses on determining an inescapable lower limit.
Let us also stress again that we are assuming spherical symmetry and that the diffusion is occurring along magnetic field lines that are radially directed. If the magnetic field configuration of a real target, including all these contributions, has a different symmetry, our description has to be accordingly modified.

We model the cascading of turbulence and the advection with Alfven waves, and the turbulence power spectrum evolves according to \cite{1993ApJ...413..619J,1995ApJ...452..912M}
\begin{equation}
\label{eq:Weq}
    \frac{\partial W}{\partial t}
    = \frac{\partial }{\partial k}\left[D_{kk}(W)\frac{\partial W}{\partial k}\right]
    - \frac{1}{r^2}\frac{\partial }{\partial r}(r^2 v_{A}W) 
    + \Gamma_{\rm CR}(n_e,k) W
     \, .
\end{equation}
The first term on the RHS describes the turbulent cascade by a diffusion in $k$, the second term accounts for advection with Alfv\'en velocity $v_A$, and the third term includes the source of turbulence given by resonant streaming instability.
We assume Kolmogorov turbulence, setting the wave number diffusion coefficient to:
$${D_{kk}(W)=c_k v_{\rm A}k^{7/2}\sqrt{W}},$$
with the numerical factor $c_k=0.052$.

The growth rate of the turbulence through streaming instability is given by \cite{Evoli:2018nmb,Jacobs:2021qvh}:
\begin{equation}
\label{eq:GammaCR}
    \Gamma_{\rm CR}= \frac{4\pi\,c\,v_A}{3\,k\,W(k)\,B_0^2/(8\pi)}\left[\beta (p)\,p^4\left| \frac{\partial f }{\partial r}\right|\right]_{p=p_{res}}
     \, ,
\end{equation}
where $B_0$ is the coherent magnetic field of the system, $4\pi\,p^2\,f\,dp=n_e\,dE$  and we consider $E\simeq p$ since the electrons are relativistic.
In Eq.~\eqref{eq:Weq} we neglected damping. There are different damping terms discussed in the literature. A nice overview is given for example in Ref. \cite{Recchia:2021vfw}. For the systems under consideration, i.e., with low stellar turbulence, the most important process would be the ion-neutral damping, due to momentum transfer or charge exchange of ions and neutrons. We model it adding a term $-\Gamma_{\rm IND} W$ to the RHS of Eq.~\eqref{eq:Weq} with $\Gamma_{\rm IND}$ as described in Ref.~\cite{Jacobs:2021qvh}.
By considering reasonable plasma densities for dwarf galaxies, we found that damping is not relevant in these systems. Thus, for simplicity, we neglect the damping term in the rest of the paper.
 
We solved the two coupled Eqs.~\eqref{eq:neq} and \eqref{eq:Weq} numerically, with a Crank-Nicolson and explicit scheme, respectively. We refer the reader to Appendix~\ref{sec:num} for details on the numerical implementation.  

\begin{figure}[ht!]
\centering
\includegraphics[width=0.5\textwidth]{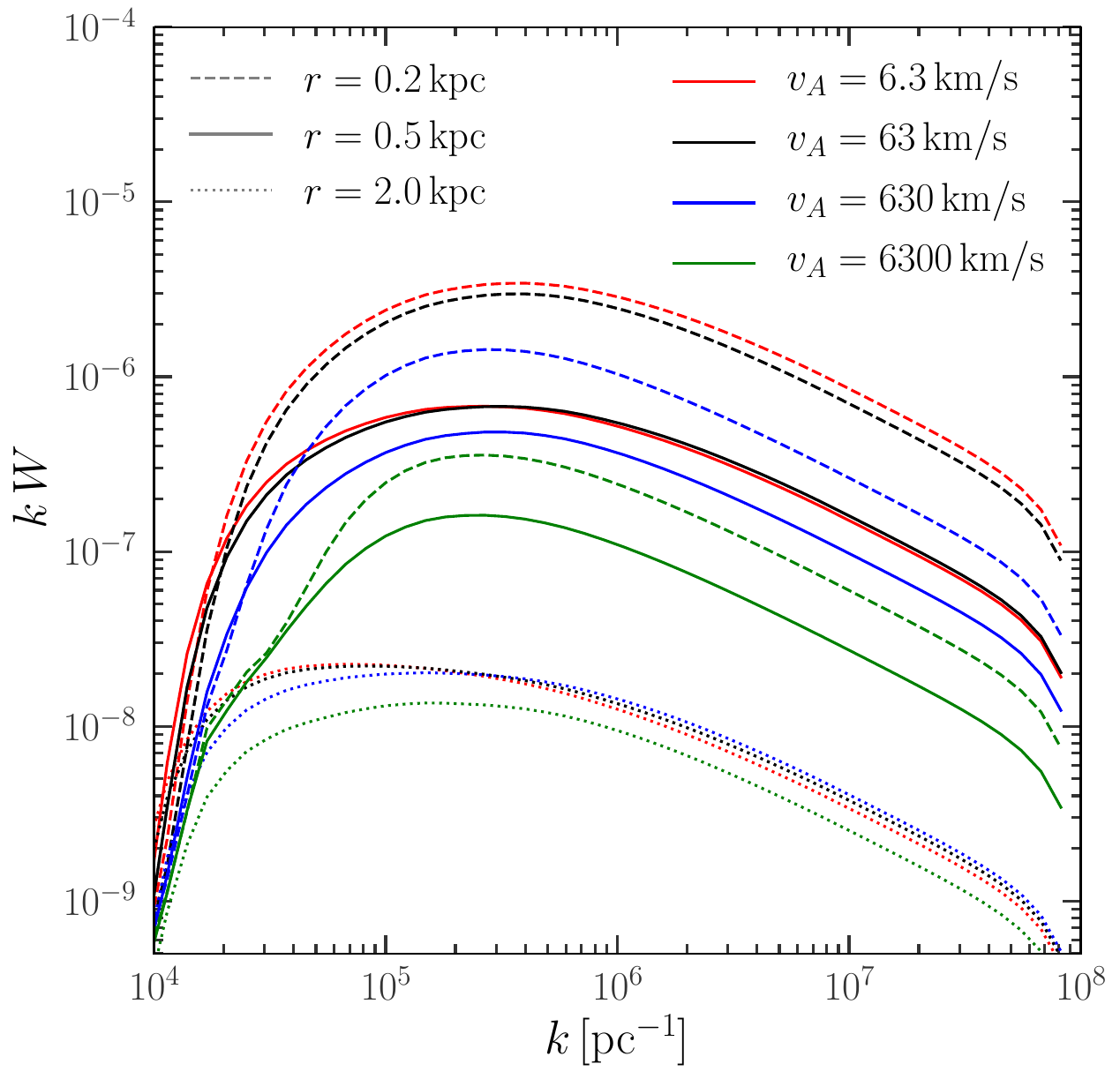}
    \caption{Power spectrum of the magnetic turbulence as a function of the wave number, for the same reference DM scenario of Fig.~\ref{fig:W_example}, but with the annihilation cross section increased by one order of magnitude, $\langle \sigma v \rangle=10^{-23}\,\rm{cm^3/s}$. We show the effect of varying the Alfv\'en speed, considering $v_A=6.3$ km/s (red), $v_A=63$ km/s (black), $v_A=630$ km/s (blue), and $v_A=6300$ km/s (green) and report the results at three different distances from the center of the system, 0.2 kpc (dashed), 0.5 kpc (solid), and 2.0 kpc (dotted). }
\label{fig:D_example2}
 \end{figure}

To understand the level of turbulence that can be provided by DM and the associated diffusion coefficient, we show some illustrative solutions in Figs.~\ref{fig:W_example} and \ref{fig:D_example2}.
 For the plots of Sections~\ref{sec:eqs} and \ref{sec:dSph}, we define a reference model with the following ingredients (reported also in Tab.~\ref{Table:ref}): $m_{\rm DM}=100$ GeV, $\langle \sigma v \rangle=10^{-24}\,\rm{cm^3/s}$, annihilation into $b\bar b$, $\rho_s=4\times 10^7\,M_\odot/{\rm kpc}^3$, $r_s=1$ kpc (the latter two numbers are chosen to have a typical DM density of dSph galaxies, see later the case of Draco), $B_0=1.0\,\mu$G and $v_A= 63$ km/s.
 In each figure, some of these parameters can be varied, and this is specified in the captions and in the following text.

\begin{table}[h]
\centering
 \begin{tabular}{||c | c | c | c | c | c | c | c ||} 
 \hline
 $B_0$ & $v_a$  & DM & $\rho_s$  & $r_s$ & $m_{\rm DM}$ & $\langle \sigma v \rangle $ & annihilation \\ [0.5ex] 
 [$\mu$G] & [km/s] & profile & $[M_\odot/\rm{kcp}^3 ]$& [kpc] & [GeV] & [$\rm{cm^3/s}$] & channel  \\ [0.5ex] 
 \hline\hline
 1 & 63  & NFW & $4\times 10^7$  & 1 & 100 & $10^{-24}$ & $b\bar b$\\ 
 \hline
\end{tabular}
\caption{Parameters for the ``reference" model used in Sections~\ref{sec:eqs} and \ref{sec:dSph}. In Figs.~\ref{fig:W_example}-\ref{fig:Dne_B0} some of the above parameters are varied, as specified in the caption of each figure and in the text.}
\label{Table:ref}
\end{table}

From both Figs.~\ref{fig:W_example} and \ref{fig:D_example2}, one can notice that the level of turbulence generated by DM increases towards the center. The main reason behind this behaviour is simply the higher DM density (and thus source term) at smaller radii. In the left panel of Fig.~\ref{fig:W_example}, we see that as $B_0$ increases, $W$ decreases. This can be understood from Eq.~\eqref{eq:GammaCR} with $B_0$ being at the denominator. In other words, the level of turbulence is set by the ratio between the turbulent magnetic field induced by DM and the coherent magnetic field, and, for a higher value of the latter, one gets the same ratio (i.e., the same turbulence) only for a higher DM source term. 

In the right panel of Fig.~\ref{fig:W_example}, we show the diffusion coefficient obtained from Eq.~\eqref{eq:dc}, for the same reference scenario, and investigating the impact of the size and normalization of the DM mass density.
First we vary $r_s$, keeping fixed $\rho_s$ (red, black and blue curves) and see that the confinement becomes more and more effective (i.e., the diffusion coefficient decreases) as the source size increases. Note that we do not consider diffusion coefficient larger than $10^{31} {\rm cm^2/s}$. The reason is related to the fact that Eq.~\eqref{eq:neq} is an effective equation, in particular it is not covariant. This means that for large $D$ (and fixed size of the diffusion region), it can describe propagation of particles faster than light. To prevent this issue, we impose $D\le 10^{31} {\rm cm^2/s}$. For such maximal value of $D$, the picture is essentially equivalent to a free escape~\cite{Regis:2014koa}, so in other words we treated all the cases providing $D> 10^{31} {\rm cm^2/s}$ as free escape.

In the right panel of Fig.~\ref{fig:W_example} we also show a case (green line) with a ``small" size of the source, $r_s=0.1$ kpc (compared to $\sim$ 1 kpc of a dwarf galaxy), but with $\rho_s$ increased by one order of magnitude with respect to the reference scenario, $\rho_s=4\times 10^8\,M_\odot/{\rm kpc}^3$. 
This shows that the effect we are describing in this work, with a focus on dwarf galaxies, can be relevant in general for dark clumps, that, on the other hand, must have large DM over-densities.

Fig.~\ref{fig:D_example2} shows that the effect of advection is marginal, for the systems under consideration. It can have an impact only for extremely large values of the Alfv\'en speed, $v_A\gtrsim 10^3$ km/s. This can be understood by estimating the time-scales associated to the three physical processes included in the equations:
\begin{equation}
    \tau_{\rm diff}\simeq 10^{15}\,{\rm s}\left(\frac{L}{{\rm kpc}}\right)^2\frac{10^{28}{\rm cm^2/s}}{D}\;,\;\tau_{\rm adv}\simeq 3\times 10^{15}\,{\rm s}\,\frac{L}{{\rm kpc}}\frac{10\,{\rm km/s}}{v_A}\;,\;\tau_{\rm loss}\simeq 4\times 10^{16}\, {\rm s}\,\frac{{\rm GeV}}{E}\left[1+0.1\left(\frac{B}{\mu{\rm G}}\right)^2\right]\,.
    \label{eq:tau}
\end{equation}
These estimates tell us that for $D\gtrsim 10^{28}{\rm cm^2/s}$, advection and energy losses are typically subdominant, and that for $D\simeq 10^{29}{\rm cm^2/s}$, which is the case for different benchmark cases shown in our plots, one needs $v_A\sim 10^3$ km/s to make advection the dominant transport mechanism. In Fig.~\ref{fig:D_example2}, we increased the annihilation cross section to $\langle \sigma v \rangle=10^{-23}\,\rm{cm^3/s}$ in order to have a picture with lower $D$, i.e., to have scenarios with more relevant impact from advection. 

\section{The case of dwarf spheroidal galaxies and synchrotron radiation}
\label{sec:dSph}

\begin{figure}[t!]
\centering
\includegraphics[width=0.45\textwidth]{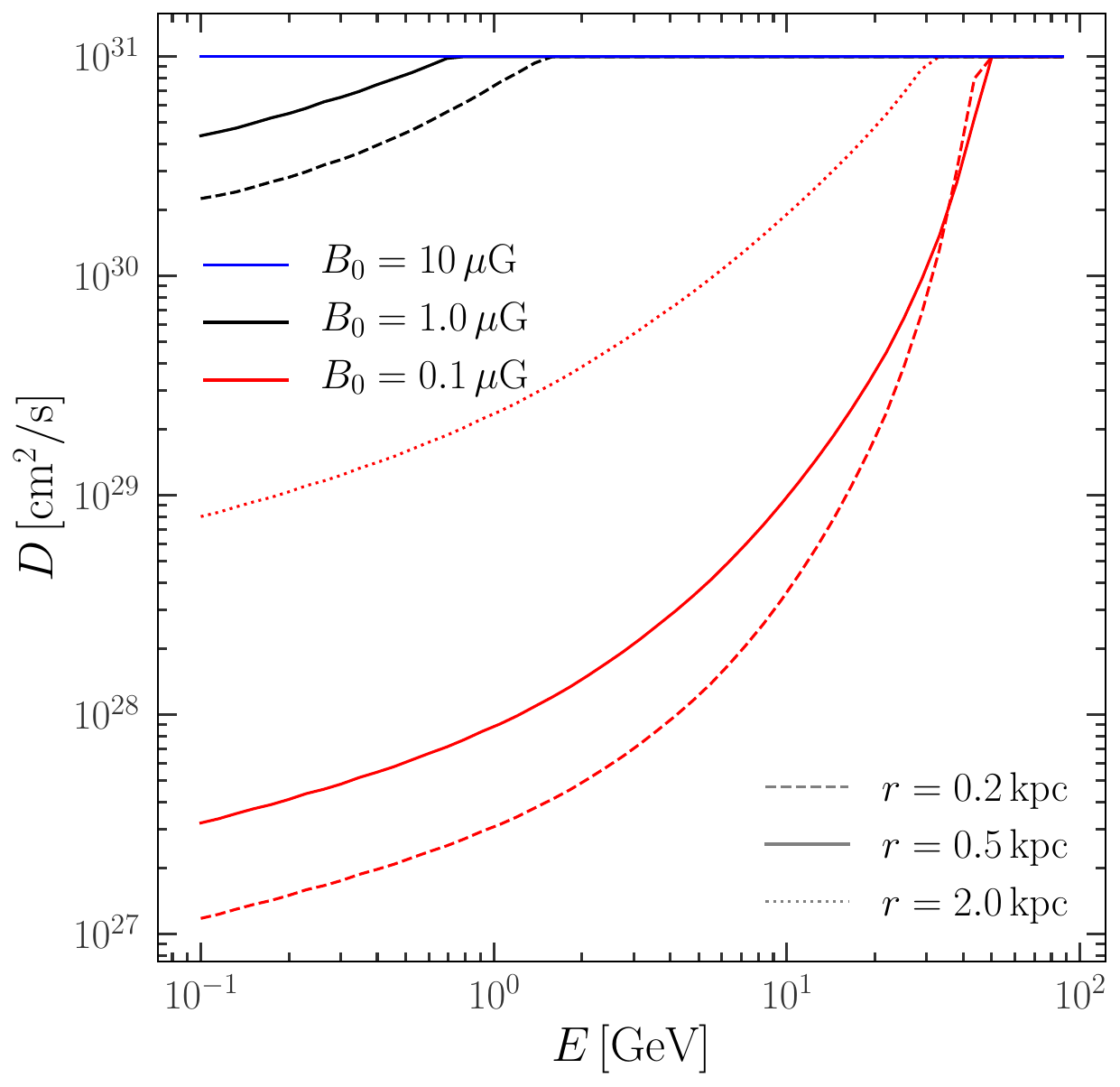}%
\hspace{0.3cm}
\includegraphics[width=0.46\textwidth]{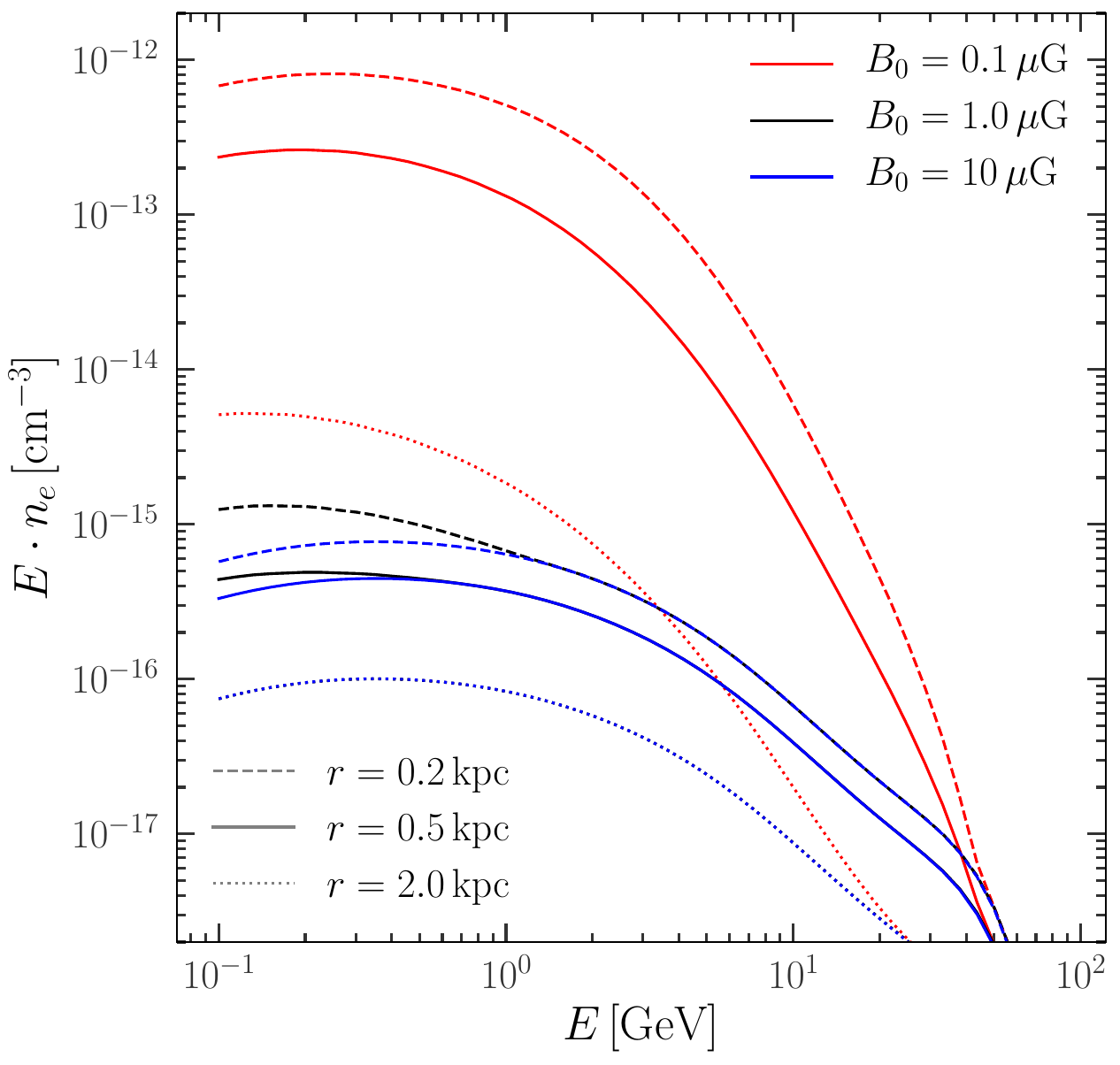}%
    \caption{Spatial diffusion coefficient (left) and equilibrium number density of electrons (right) as a function of energy for a reference scenario (detailed in the text) and investigating the dependence on the regular magnetic field. We show three different radial distances, 0.2 kpc (dashed), 0.5 kpc (solid), and 2.0 kpc (dotted), and three different magnetic field strengths $B_0=0.1\,\mu$G (red), $1\,\mu$G (black), and $10\,\mu$G (blue, where Eq.~\eqref{eq:dc} would give $D> 10^{31} {\rm cm^2/s}$). 
}
\label{fig:Dne_B0}
 \end{figure}

\begin{figure}[t!]
\includegraphics[width=0.45\textwidth]{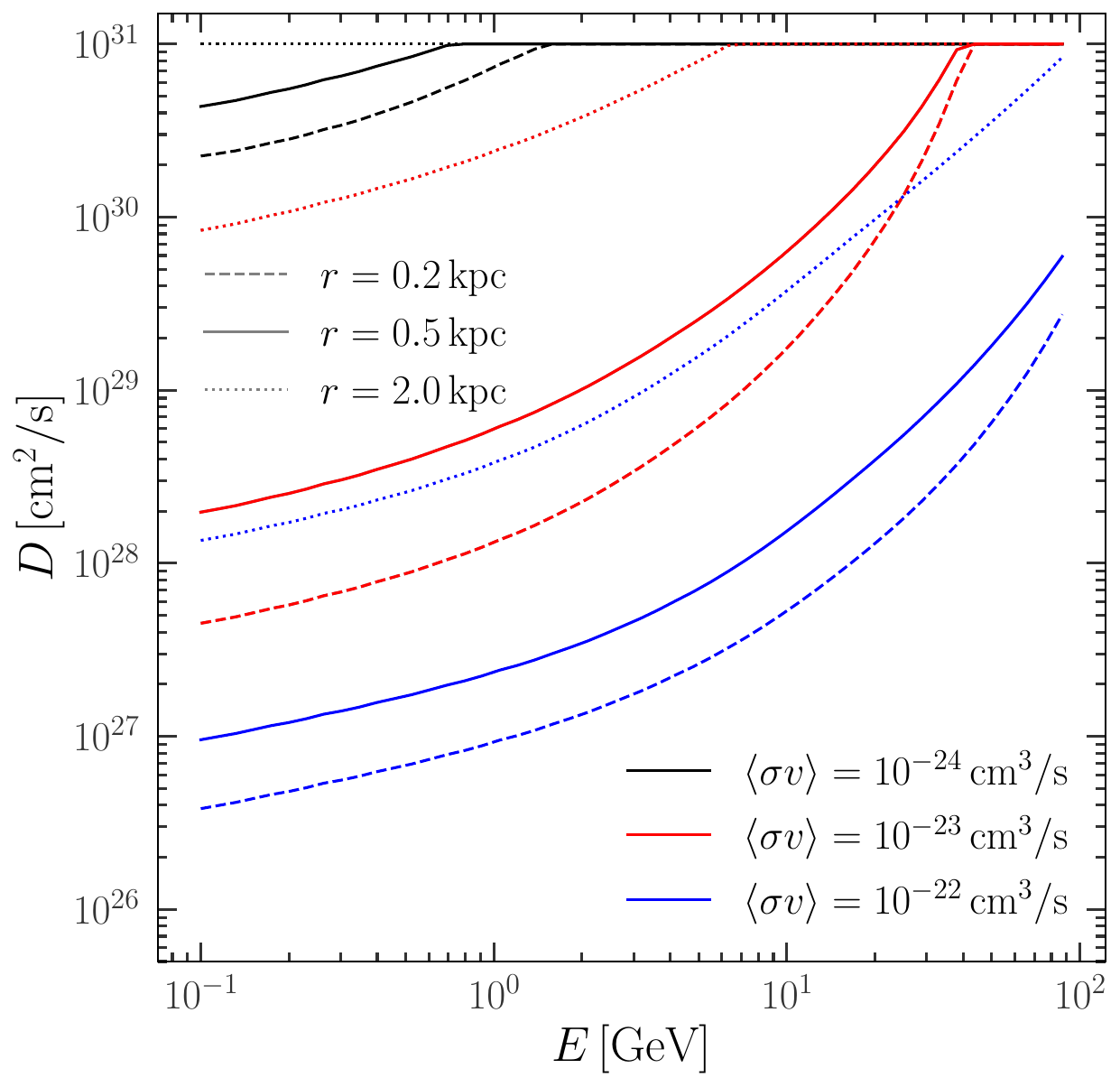}
\hspace{0.3cm}
\includegraphics[width=0.46\textwidth]{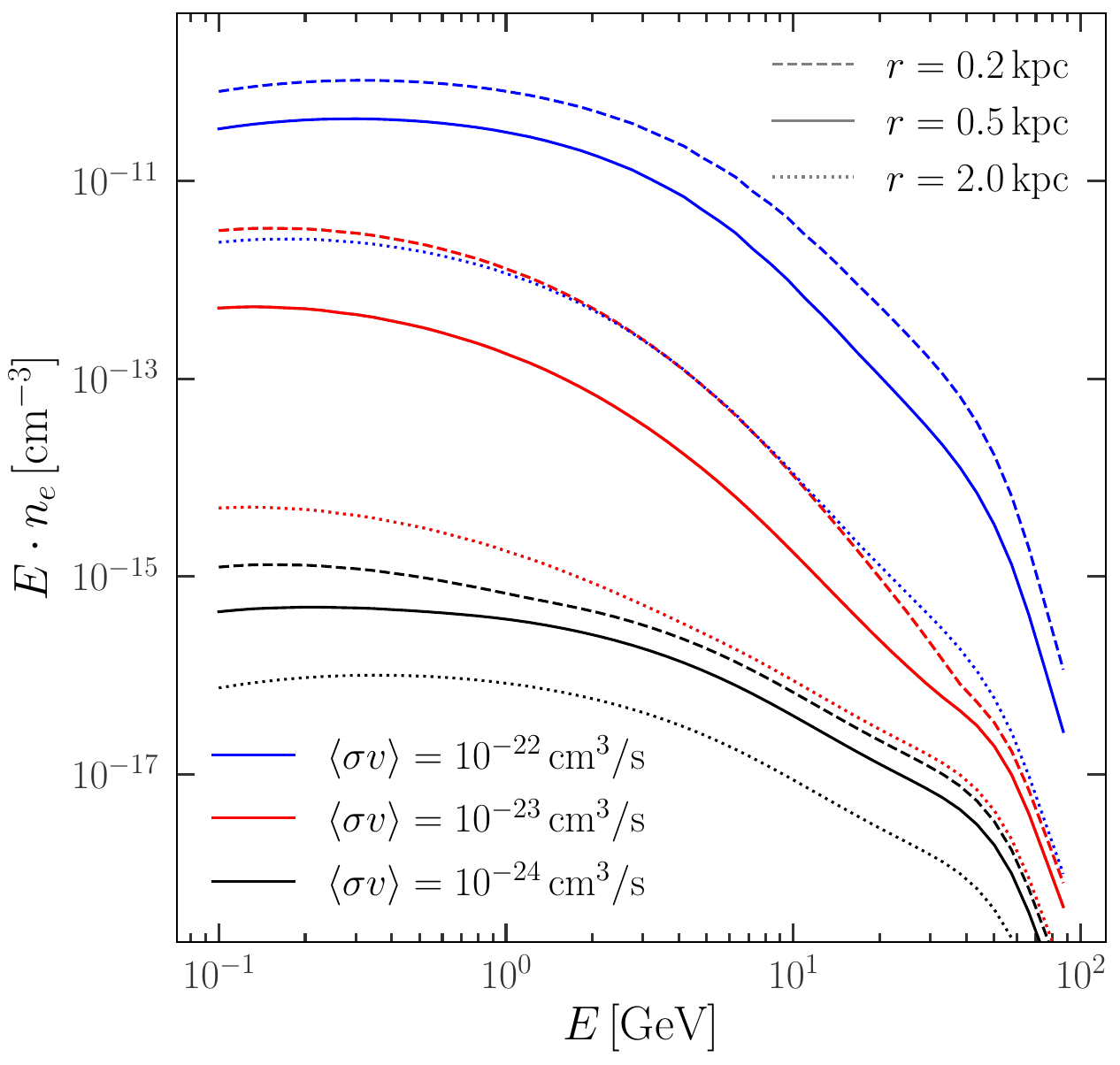}
    \caption{Spatial diffusion coefficient (left) and equilibrium number density of electrons (right) as a function of energy for a reference scenario (detailed in the text) and investigating the dependence on the annihilation cross section. We show three different radial distances, 0.2 kpc (dashed), 0.5 kpc (solid), and 2.0 kpc (dotted), and three different annihilation cross sections $\langle \sigma v \rangle=10^{-24}\,\rm{cm^3/s}$ (black), $\langle \sigma v \rangle=10^{-23}\,\rm{cm^3/s}$ (red), and $\langle \sigma v \rangle=10^{-22}\,\rm{cm^3/s}$ (blue). }
\label{fig:Dne_sv}
 \end{figure}

In this Section, we set the DM profile to the case of a prototypical dSph galaxy. In particular, we consider again an NFW profile with $\rho_s=4\times 10^7\,M_\odot/{\rm kpc}^3$ and $r_s=1$ kpc, that are the best-fit values found in the case of Draco, see Appendix~\ref{sec:jeans}. Here Draco is taken as a benchmark dSph, and considering other classical dSph galaxies would produce very similar plots.
In Figs.~\ref{fig:Dne_B0} and \ref{fig:Dne_sv} we show again solutions of Eqs.~\eqref{eq:neq} and \eqref{eq:Weq} (the latter reported in terms of $D$ instead of $W$ using Eq.~\eqref{eq:dc}).

First, we explore the dependence on the strength of the coherent magnetic field. We assume this magnetic component to be present independently from DM. Plausible estimates point towards a strength at the level of $\mu$G or fraction of $\mu$G, see, e.g., discussion in~\citep{Regis:2014koa}. In Fig.~\ref{fig:Dne_B0}, we take $\langle \sigma v \rangle=10^{-24}\,\rm{cm^3/s}$ (and all the other parameters as in the reference model of Table~\ref{Table:ref}), and see that for $B_0<1\,\mu$G, the self-confinement is important. Indeed for $B_0\sim$ fraction of $\mu$G, we get low values of $D$, and electron densities that are orders of magnitude larger than in the free escape case (as it can be seen by comparing red and blue curves in the right panel). 

In Fig.~\ref{fig:Dne_sv}, we set $B_0=1\,\mu$G and explore the dependence on the annihilation cross section. It is clear that for larger $\langle \sigma v \rangle$ the source term is larger, and so the generation of turbulence is more effective, leading to a lower diffusion coefficient and larger electron density. It is important to note that the effect deviates from a simple linear scaling. There are two saturation effects: for too low level of injection from DM, one always gets a ``free-escape" scenario, whilst on the contrary for very large injection one ends up in a scenario where the residence time is mostly determined by advection (see discussion of Eq.~\eqref{eq:tau}), which is independent from $\langle \sigma v \rangle$.

From the determination of the $e^+-e^-$ number density $n_e$ that we just discussed, we can now compute the synchrotron emissivity at a given frequency $\nu$ by folding $n_e$ with the total radiative emission power $P_{syn}$ \citep{Rybicki:1979}:
\be
j_{syn}(\nu,r)=\int dE\,P_{syn}(r,E,\nu)\, n_e(r,E)\;.
\label{eq:jsynch}
\ee
The unpolarized synchrotron power depends on the total magnetic field $B$, i.e., on the sum of the coherent (that we call $B_0$) plus random (that we call $\delta B$) magnetic fields. We compute the latter from $\delta B=B_0\,\sqrt{\int dk\,W(k)}$. For models in the ball-park of the bounds we are going to derive in the next Section, $\delta B \ll B_0$, and thus $B\simeq B_0$, which is reassuring, since we are working under the quasi-linear approximation. 

It is well known that the synchrotron power roughly scales with $B^2$ with a cutoff for $B\lesssim \nu/\rm{GHz}\,(15\,\rm{GeV}/E)^2$. Since $B$ enters in the determination of $n_e$, the dependence of the emissivity from the magnetic field is not trivial (and depends on $m_{\rm DM}$, $\langle \sigma v \rangle$ and on the annihilation channel).

\begin{figure}[ht!]
\includegraphics[width=0.48\textwidth,trim={0.5cm 1.05cm 3.5cm 3.5cm},clip]{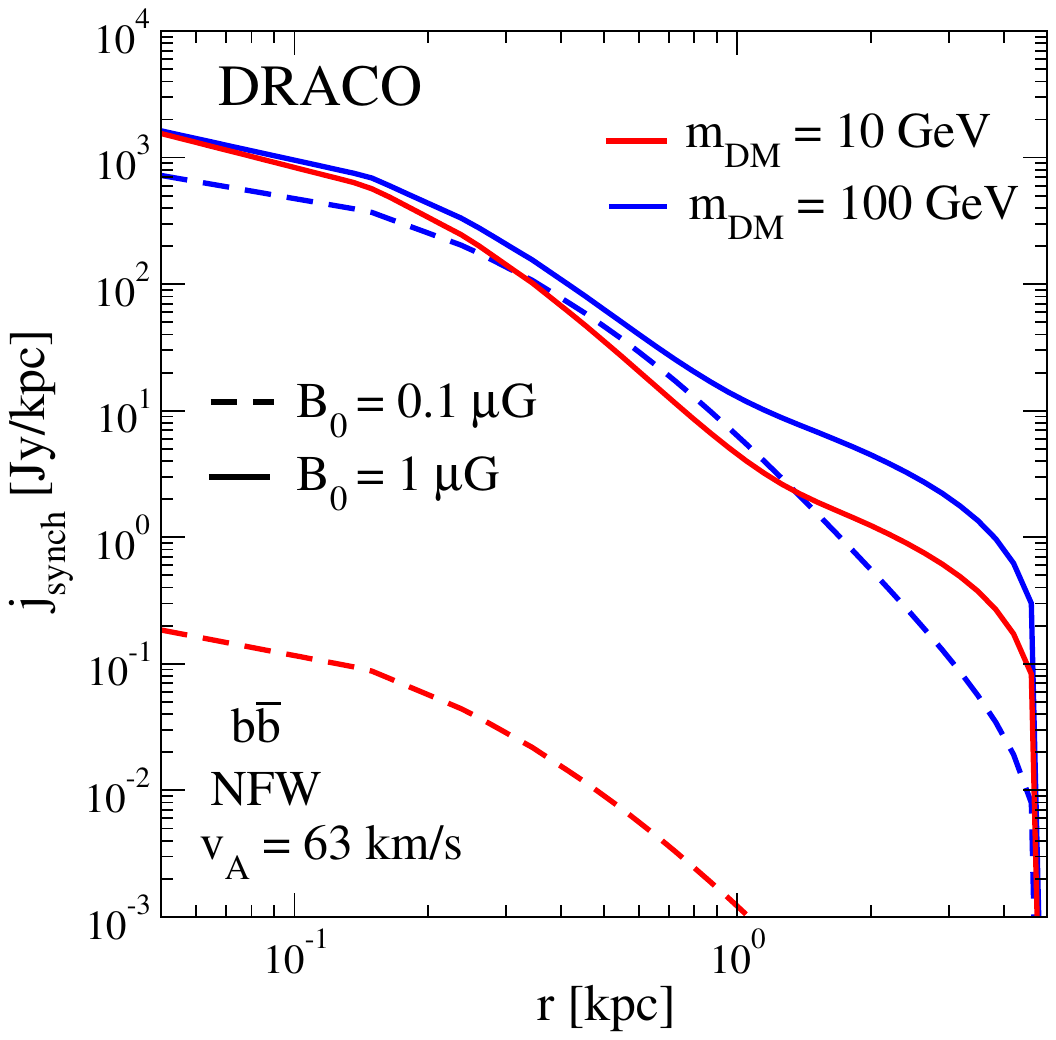}\hspace{0.04\textwidth}%
\includegraphics[width=0.48\textwidth,trim={0.5cm 1.05cm 3.5cm 3.5cm},clip]{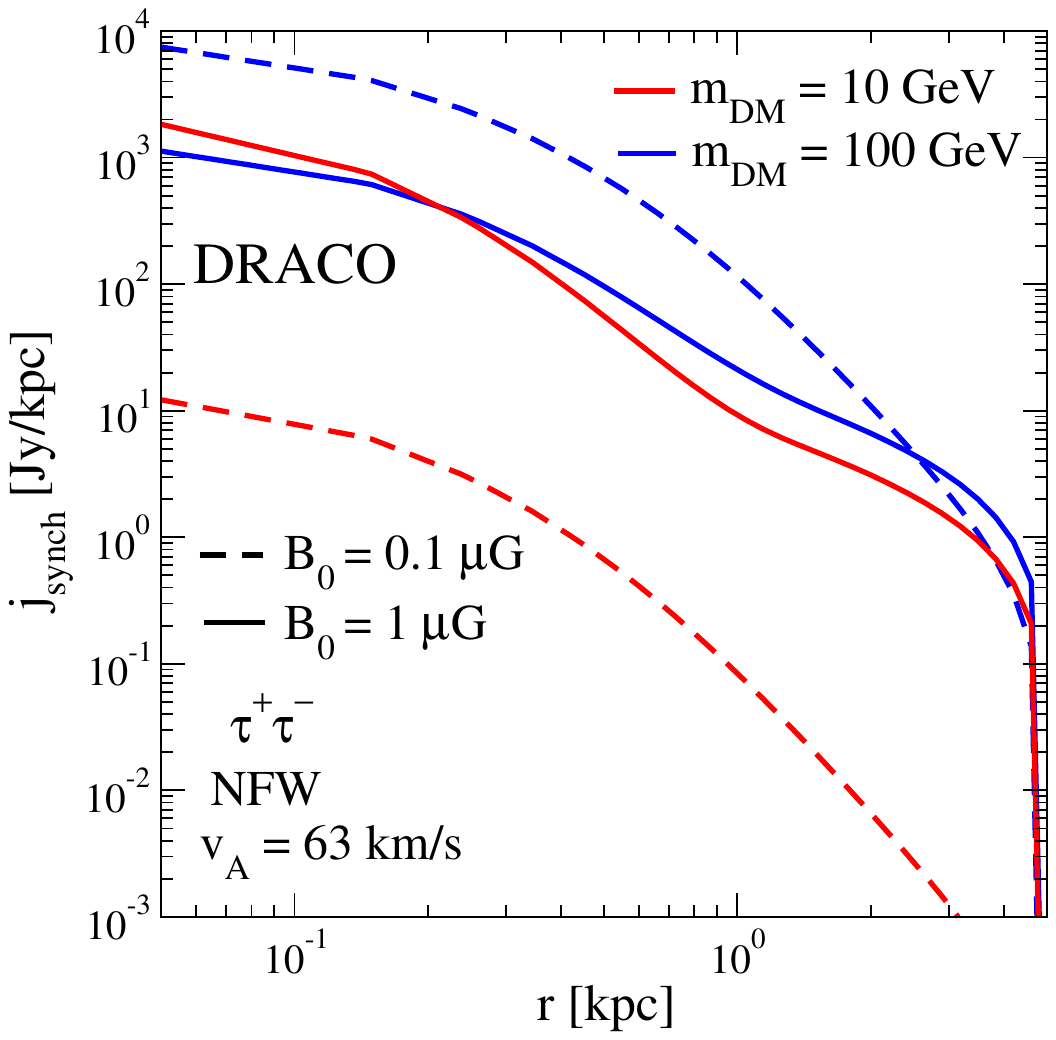}
    \caption{Emissivity at 650 MHz as a function of the radial distance from the dSph center. In the left panel we consider the $b\bar b$ annihilation channel, and set the annihilation cross section to $\langle \sigma v \rangle = 3.3\times 10^{-25}\,\rm{cm^3/s}$ ($1.5\times 10^{-24}\,\rm{cm^3/s}$) for $m_{\rm DM}=10$ GeV (100 GeV). On the right panel we show the case of annihilation into $\tau^+\tau^-$ with $\langle \sigma v \rangle =1.4\times 10^{-25}\,\rm{cm^3/s}$ ($2.9\times 10^{-24}\,\rm{cm^3/s}$) and $m_{\rm DM}=10$ GeV (100 GeV).  }
\label{fig:jsynch}
 \end{figure}

We show some illustrative cases in Fig.~\ref{fig:jsynch}, taking again a typical dSph, Draco, and a radio frequency of 650 MHz. We vary the magnetic field strength between 0.1 and 1 $\mu$G, and consider two DM masses, 10 and 100 GeV, and two annihilation channels, $b\bar b$ and $\tau^+\tau^-$. We set the cross section so to have non-negligible confinement (more precisely we choose the value corresponding to the limit, computed in the next Section, for $B_0=1\,\mu$G). 

We see that for $m_{\rm DM}=100$ GeV, i.e., for scenarios efficiently injecting electrons at energies far from the cutoff of the synchrotron power, an increase of $B_0$ from 0.1 to 1 $\mu$G does not lead to significantly larger fluxes, or even to lower emissivity, see the $\tau^+\tau^-$ case. Namely, the reduction in $n_e$ for larger $B_0$, due to a smaller confinement time, is compensating the increase in the synchrotron power, and the dependence of the signal on the $B_0$ strength is mild. On the other hand, for lower DM masses, the scaling with $B_0$ can be dramatic since the peak of the distribution of the injected $e^+-e^-$ is now at an energy lower than the one corresponding to the cutoff in the synchrotron power.

\section{Bounds from radio observations of the Draco dSph galaxy}
\label{sec:radio}
Our final goal is to derive bounds on WIMP DM by comparing the radio emission induced by DM in dSph galaxies with observations.

The flux density measured by a telescope can be estimated as
\be
S_{th}(\nu,\theta_0) =\int d \phi\,d\theta\,\sin\,\theta\,\mathcal{G}(\theta,\phi,\theta_0)\int ds\,\frac{j_{syn}(\nu,r(s,\theta,\phi))}{4\pi}\;,
\label{eq:Isynch}
\ee
where $s$ labels the coordinate along the line of sight, $\phi$ and $\theta$ are the angular coordinates, $\theta_0$ is the direction of observation, and $\mathcal{G}$ gives synthesized beam of the telescope.

In the evaluation of Eq.~\eqref{eq:Isynch}, we need to specify five ingredients, two of them related to the WIMP microscopic properties, i.e., the particle mass $m_{\rm DM}$ and the annihilation rate $\langle \sigma v \rangle$, and three related to the dSph, namely, the DM spatial distribution, the regular magnetic field, and the plasma density (the latter to provide the Alfven velocity, see above).
We derive bounds in the plane $\langle \sigma v \rangle$ versus $m_{\rm DM}$.

We constrain the DM profile $\rho$ via Jeans analysis and using the dispersion velocities of the dSph stellar component. In particular, we consider the parametric form of the NFW profile~\cite{Navarro:1995iw}, and derive the likelihood $\mathcal{L}_{Jeans}$ associated to the parameters $(\rho_s,r_s)$ describing the distribution. For more details, see Appendix~\ref{sec:jeans}, where we also consider the case of a cored profile described by the Burkert parametrization.

For what concerns the regular magnetic field and plasma density, we cannot proceed with a data-driven determination since only upper limits can be found in the literature. Thus we bracket the uncertainty by considering two different scenarios for each quantity, one where we assume values smaller than in the host galaxy (i.e., our Milky Way) but not too far-away, taking $B_0=1\,\mu$G and $n_g=10^{-3}\,{\rm cm}^{-3}$, and one where instead we take a significantly smaller estimate but still above the cosmological values, i.e., $B_0=0.1\,\mu$G and $n_g=10^{-5}\,{\rm cm}^{-3}$.

With this modeling at hand we can now evaluate $S_{th}$ for a given dSph. In particular, we compute it for the Draco dSph and compare to recent data collected with the upgraded Giant Metrewave Radio Telescope (uGMRT) at 550-750 MHz. For more details on the observations, see Appendix~\ref{sec:obs}.

Assuming a Gaussian likelihood, we can compare expected diffuse signals and observed data through~\cite{Regis:2014koa}:
\be 
\mathcal{\tilde{L}}_{diff}=e^{-\chi^2/2} \;\;\; {\rm with} \;\;\; \chi^2=\frac{1}{N_{pix}^{FWHM}}\sum_{i=1}^{N_{pix}} \left(\frac{S_{th}^i-S_{obs}^i}{\sigma_{rms}^i}\right)^2\;,
\label{eq:like}
\ee
where $i$ denotes the pixel in the Draco radio image, $S_{obs}^i$ is the observed flux density, $\sigma_{rms}^i$ is the r.m.s. error, $N_{pix}$ is the total number of pixels in the area under investigations, that is chosen to be a circle of $30^\prime$ of radius, and $N_{pix}^{FWHM}$ is the number of pixels within the uGMRT synthesized beam.

We add to the theoretical term $S_{th}$ a spatially flat term $S_{flat}$ that is included in the fit to account for a possible offset in the zero-level calibration of the map, and then we define a likelihood $\mathcal{L}_{diff}$ which depends only on the DM parameters by profiling out $S_{flat}$ from the likelihood in Eq.~\eqref{eq:like}.

Then, combining the Jeans analysis and uGMRT data, we can define, at any given mass $m_{\rm DM}$, a global likelihood:
\be 
\mathcal{L}(\langle \sigma v \rangle,\rho_s,r_s)=\mathcal{L}_{diff}(\langle \sigma v \rangle,\rho_s,r_s)\times \mathcal{L}_{Jeans}(\rho_s,r_s) \;.
\label{eq:likeall}
\ee

To derive the bounds, we assume that $\lambda_c\langle \sigma v \rangle=-2\ln[\mathcal{L}(\langle \sigma v \rangle,\rho_s^{\,lbf}, r_s^{\,lbf})/\mathcal{L}(\langle \sigma v \rangle^{b.f.},\rho_s^{\,gbf},r_s^{\,gbf})]$ follows a $\chi^2$-distribution with one d.o.f. and with one-sided probability given by $P=\int^{\infty}_{\sqrt{\lambda_c}}d\chi\,e^{-\chi^2/2}/\sqrt{2\,\pi}$, where $\langle \sigma v \rangle^{b.f.}$ denotes the best-fit value for the annihilation rate at that specific WIMP mass. The superscript $gbf$ indicates the global best-fit (i.e., taking $\langle \sigma v \rangle=\langle \sigma v \rangle^{b.f.})$), whilst $lbf$ denotes the best-fit of $\rho_s$ and $r_s$ for that given $\langle \sigma v \rangle$. 
The 95\% C.L. upper limit on $\langle \sigma v \rangle$ at mass $m_{\rm DM}$ is obtained from $\lambda_c=2.71$.

\begin{figure}[b!] 
\includegraphics[width=0.49\textwidth,trim={0.4cm 0.79cm 3.0cm 3.5cm},clip]{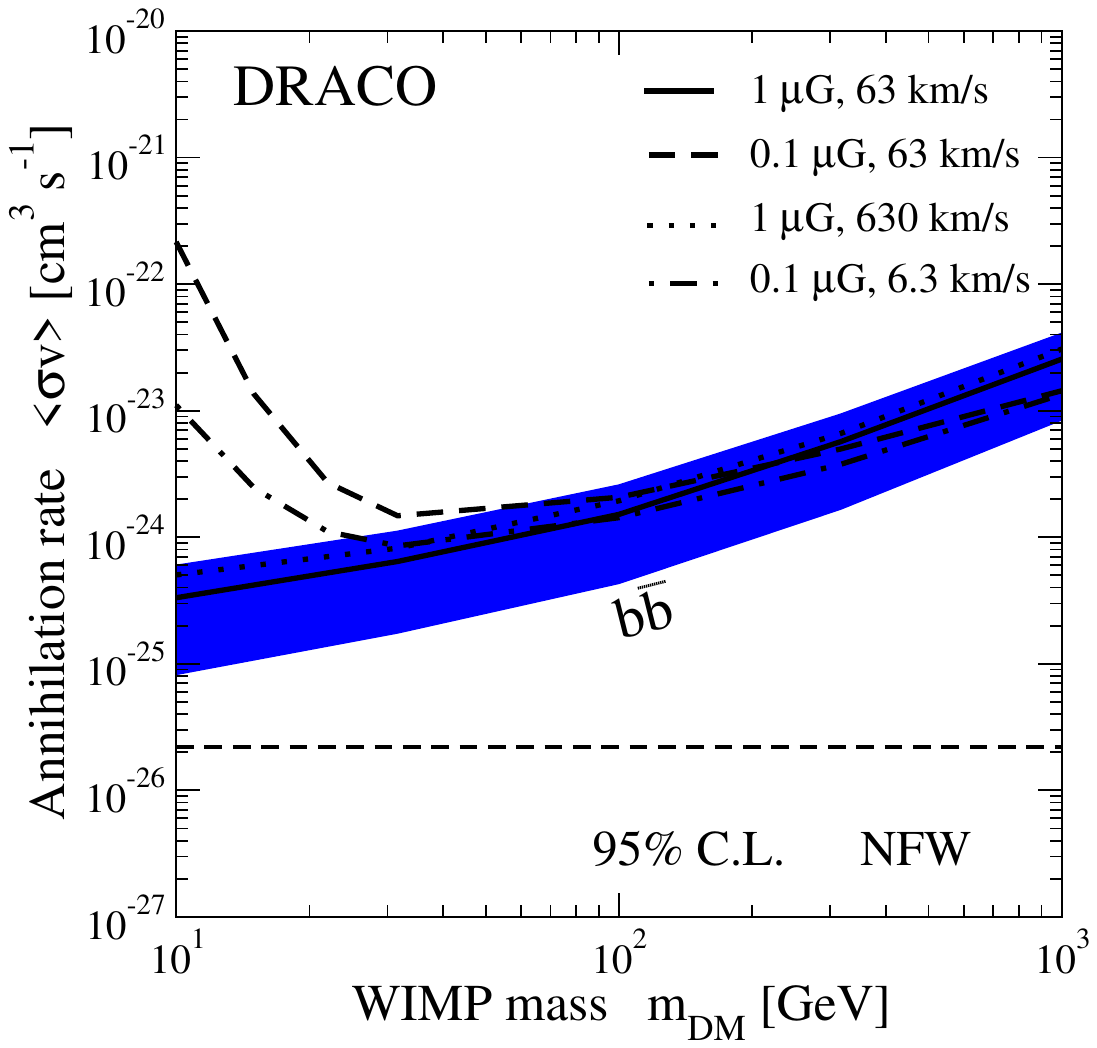}\hspace{0.015\textwidth}%
\includegraphics[width=0.49\textwidth,trim={0.4cm 0.79cm 3.0cm 3.5cm},clip]{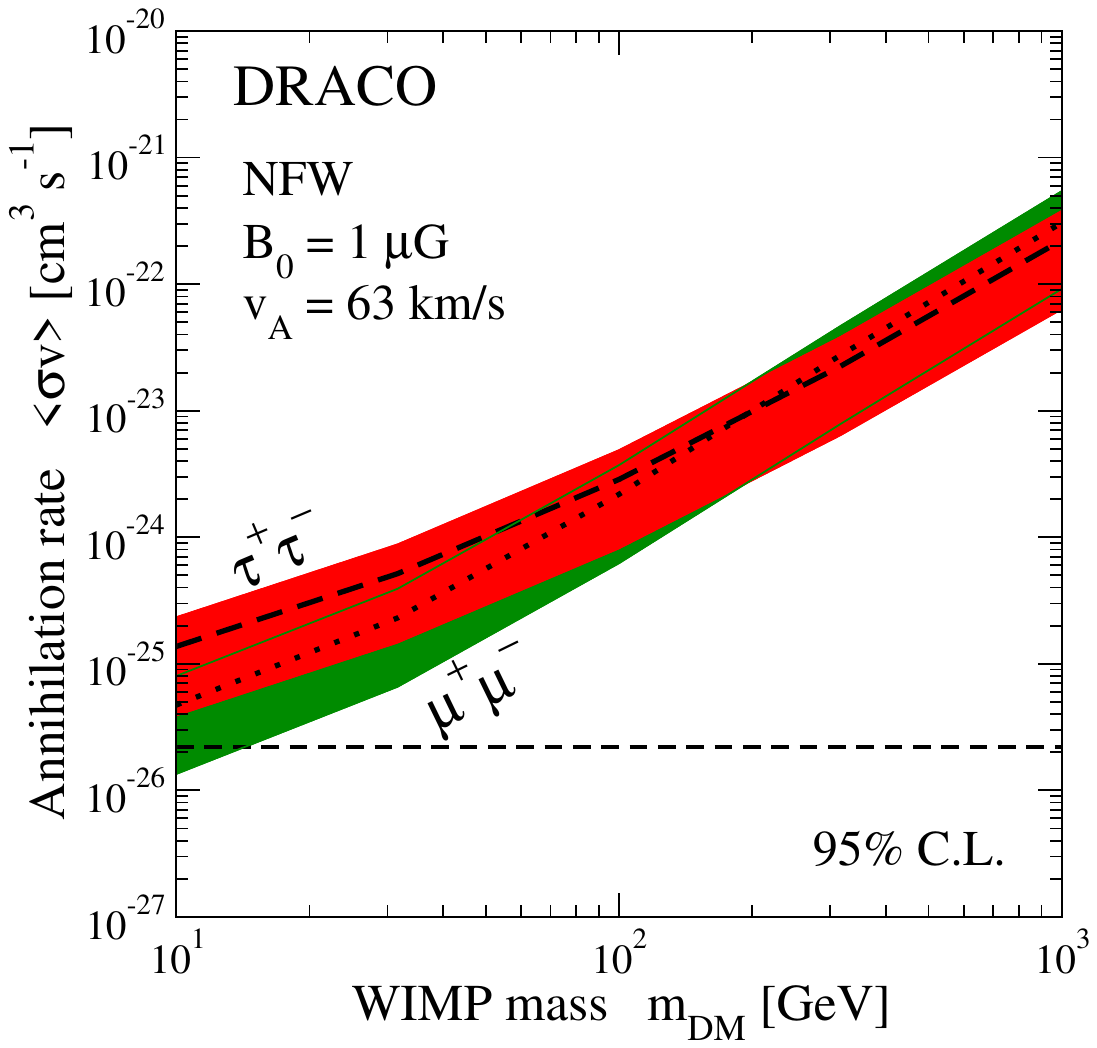}
    \caption{Bounds at 95\% C.L. on the annihilation rate as a function of the DM mass. Our reference scenario has $B_0=1\,\mu$G and $n_p=10^{-3}\,{\rm cm^{-3}}$ (which implies $v_A\simeq 63$ km/s).
    The colored bands show the impact of the uncertainty associated to the determination of the DM density, assumed to follow an NFW profile. Blue is for $b\bar b$, red for $\tau^+\tau^-$ and green for $\mu^+\mu^-$. In the left panel, we show also alternative scenarios, where $B_0=0.1\,\mu$G and/or $n_p=10^{-5}\,{\rm cm^{-3}}$. }
\label{fig:bounds}
 \end{figure}

Results are shown in Fig.~\ref{fig:bounds} with black lines for three different annihilation channels, $b\bar b$ (left), $\tau^+\tau^-$ (right, dashed) and $\mu^+\mu^-$ (right, dotted). The reference scenario assumes $B_0=1\,\mu$G and $n_p=10^{-3}\,{\rm cm}^{-3}$ (i.e., $v_A=63$ km/s). In the left panel, we show the effect of varying the magnetic field and advection velocity, by considering $B_0=0.1\,\mu$G and $n_p=10^{-5}\,{\rm cm}^{-3}$ (we focus on the $b\bar b$ channel for the sake of brevity). To understand the values of $v_A$, we remind that $v_A\simeq 63\,{\rm km/s}\,\sqrt{10^{-3}{\rm cm^{-3}}/n_p}\,B_0/\mu{\rm G}$.

The left panel shows the main achievement of this work. We devised a self-consistent method to derive bounds on WIMP DM from radio observations of dSph. While in the current state-of-the-art one deals with several orders of magnitude of uncertainty for the derived bounds (see, e.g., Fig. 5 in \citep{Regis:2014tga}), which are due to the unknown description of the dSph interstellar medium, in this work we show there is an ``irreducible'' bound, that is only a factor $\mathcal{O}(1)$ uncertain (for masses above 20-30 GeV).

The two ingredients we need to include in the model (and that are not data-driven, like on the contrary the parameters of the DM profile) are the advection and the regular magnetic field. The impact of advection is limited. The value of $B_0$, instead, significantly affects both the synchrotron power and the confinement time, but the effects on the signal go in the opposite directions, partially canceling each other. Indeed, the larger is $B_0$ the larger is $P_{syn}$ (thus enhancing the signal), but also the larger is $D$ (which depletes the signal), as we already discussed in the previous Section. This implies that the different scenarios differ only by a factor $\mathcal{O}(1)$ in the expected signal and so in the bound.

The only exception to this argument is for scenarios with low $B_0$, low $m_{\rm DM}$ and soft channels of annihilation.
Indeed, in this case the peak of the synchrotron power corresponds to an energy well above the peak of the $e^+-e^-$ spectrum induced by WIMP annihilations. We show it in the case of $b\bar b$, where this effect is maximal, being the spectrum of emission quite soft, whilst for leptonic channels the spectrum is harder and the effect is more limited, also at low masses.

For a fair comparison of Fig.~\ref{fig:bounds} with bounds in other analyses of Draco, let us mention that the determination of the DM profile described in Appendix~\ref{sec:jeans} leads to a J-factor at $0.5^\circ$ which is in agreement with recent estimates \cite{Petac:2018gue,Alvarez:2020cmw}, but a factor of $\sim 3$ lower than in the computation of \citep{Bonnivard:2015xpq}, which has been widely used in the literature.
To understand the impact of the uncertainty in the determination of the DM profile, we include the colored bands in Fig.~\ref{fig:bounds}. They show bounds derived with a procedure that is slightly different from the one outlined above, namely, the band reports the bounds obtained by taking all the $\rho_s$ and $r_s$ value within their 95\% C.L. region (derived from the Jeans analysis).

\section{Conclusions}
\label{sec:conc}
In this work we investigated the generation of magnetic turbulence due to the injection of electrons and positrons from DM annihilation.
In objects where the star formation rate is significant, this mechanism is typically subdominant with respect to the magnetic irregularities given by, e.g., supernova explosions. On the other hand, we found that in ``dark" and DM dense systems, it can be the driving mechanism, and it can lead to a non-negligible self-confinement of the electrons and positrons.

We quantitatively assessed the effect by solving numerically a system of two coupled differential equations that describe the evolution of the electron number density and of the turbulence. 

As a concrete application, we derived solutions in the case of dSph galaxies, which are one of the prime targets in DM indirect searches. After determining the equilibrium density of electrons and positrons, we computed the associated synchrotron radiation. This emission peaks in the radio regime and the search for faint, diffuse radio sources is expected to significantly progress in the forthcoming years thanks to SKAO and its precursors.

Our findings suggest that the self-generated turbulence implies a non-negligible lower limit to the residence time of the electrons in dSphs. The dependence of this effect on the strength of the coherent magnetic field $B_0$ partially cancels the dependence of the synchrotron power on $B_0$, implying low uncertainty in the predicted DM-induced synchrotron emission. This allows to derive robust bounds, something that represents a significant improvement with respect to the current state-of-the-art where bounds are instead heavily dependent on the magnetic assumptions (see, e.g., discussion in \citep{Regis:2014tga}).

Finally, we used data of the Draco dSph collected at the uGMRT at 550-750 MHz, and compared them to the expected DM synchrotron emission computed within the self-confinement framework.
We found that the proposed approach can already allow to set robust and competitive bounds on WIMP DM, summarized in Fig.~\ref{fig:bounds}. 
Near future radio observations with the SKAO and its precursors have the capability to significantly further push down the derived constraints.

\section*{Acknowledgements}
We would like to thank C. Evoli and S. Recchia for useful discussions, and the Institute for Fundamental Physics of the Universe (IFPU), Trieste, where the early phase of this work was carried out, for the hospitality and support.

MR and JR acknowledge support from the project ``Theoretical Astroparticle Physics (TAsP)'' funded by the INFN, from `Departments of Excellence 2018-2022' grant awarded by the Italian Ministry of Education, University and Research (\textsc{miur}) L.\ 232/2016 and from the research grant `From Darklight to Dark Matter: understanding the galaxy/matter  connection to measure the Universe' No.\ 20179P3PKJ funded by \textsc{miur}.
MK is supported by the Swedish Research Council under contracts 2019-05135 and 2022-04283 and the European Research Council under grant 742104.
PU is supported by the research grant ``The Dark Universe: A Synergic Multi-messenger Approach" number 2017X7X85K under the program PRIN 2017 funded by the The Italian Ministry of Education, University and Research (MIUR), and by the European Union's Horizon 2020 research and innovation program under the Marie Sklodowska-Curie grant agreement No 860881-HIDDeN. 

\bibliography{main}

\newpage
\appendix
\section{Numerical implementation}
\label{sec:num}
In this Appendix, we report details on the numerical solution of Eqs.~\eqref{eq:neq} and \eqref{eq:Weq}.
\subsection{Electron/positron density}
Equation \eqref{eq:neq} has been finite-differenced by means of the Crank-Nicolson scheme. It has built on the case without advection, described in the Appendix of \citep{Regis:2014koa} (see also \cite{Vollmann:2020} for a semi-analytical solution). 
The full discretized equation is given by: 
\begin{align}
    \frac{\ia{n}{t+1}{i}{j} - \ia{n}{t}{i}{j}}{\Delta t} 
  &= \frac{1}{2}\SQB{\ia{L}{t}{i}{j} + \ia{L}{t+1}{i}{j}}
  \label{eq:nediscr}
\end{align}
where we defined $n(t_t, r_i, \tilde{E}_j) = \ia{n}{t}{i}{j}$ and consider a linear grid in space $r (r_i) = \io{r}{i}$ and a logarithmic grid in energy $\tilde E_j = \log\RB{ E_j}$. 
The operators in Eq.~\eqref{eq:nediscr} are given by (where each line corresponds to each of the five different terms on the RHS of Eq.~\eqref{eq:neq}):
\begin{align}
  \ia{L}{t}{i}{j} 
  = &  \frac{1}{r_i^2 (\Delta r)^2}   
  \left[ 
     \RB{ \io{r}{i+\half} }^{2} \is{D}{i+\half}{j} 
     \RB{ \ia{n}{t}{i+1}{j}-\ia{n}{t}{i}{j} } 
  \right. \\
  &\qquad\qquad\qquad\qquad- 
  \left. 
     \RB{ \io{r}{i-\half} }^{2} \is{D}{i-\half}{j}  
     \RB{ \ia{n}{t}{i}{j}-\ia{n}{t}{i-1}{j} } 
  \right] \\
  & - v_{A} \SQB{
         \frac{2}{r_i} \ia{n}{t}{i}{j}
       + \frac{1}{\Delta r} \RB{ \ia{n}{t}{i}{j} - \ia{n}{t}{i-1}{j} }
       } \\
  & -\frac{1}{E_j \Delta\tilde E} \SQB{ 
      \rb{\frac{dE}{dt}(E_{j+1})} \ia{n}{t}{i}{j+1} 
    - \rb{\frac{dE}{dt}(E_{j  })} \ia{n}{t}{i}{j  }  
  }  \\
  & + \ia{\RB{ q^{(\rm CR)} }}{t}{i}{j}. 
\end{align}
having defined $r_{i\pm\half} = (r_i + r_{i\pm1})/2$.

As mentioned, we choose a linear grid in $r$. This is because it gives a faster convergence than a log grid. We checked the results are however independent from this choice. We set $L=5\,r_s$ (where $r_s$ is the scale radius of the DM profile) and define
\begin{align*}
    \Delta_r &= \frac{L}{n_r-0.5} \qquad,\qquad
    r_i = \RB{ i + \frac{1}{2}} \Delta_r
    \quad \bigg| i \in \lbrace 0,1,...,n_r-1 \rbrace 
\end{align*}

We impose the following boundary conditions:
\begin{align}
  \partial_r n \RB{ r=0, E, t} &= 0    
  \qquad \qquad \qquad \longrightarrow &  \ia{n}{t}{-1}{j} &= \ia{n}{t}{0}{j} \\
  n \RB{ r=L, E, t} &= 0 
  \qquad \qquad \qquad \longrightarrow & \ia{n}{t}{n_r-1}{j} &= 0 \\
  n \RB{ r, E=E_{\rm min}, t} &= 0 
  \qquad \qquad \qquad \longrightarrow & \ia{n}{t}{i}{0} &= 0 \\
  n \RB{ r, E=E_{\rm max}, t} &= \text{continuous} 
  \qquad \longrightarrow & \ia{\D n}{t}{i}{n_E-1} &= \ia{\D n}{t}{i}{n_E-2}
\end{align}

Note that we include a fictitious value in $r$ below grid, $r_{-1}=-\Delta_r/2$. In this way the boundary condition is exactly at $r=0$. This is one of the motivations for choosing a linear grid since with a log grid one can only impose the boundary condition at a small $r\ne 0$.

\subsection{Magnetic turbulence}
Equation \eqref{eq:Weq} is solved using the explicit scheme.\\
First we define $Q^{(W)}=\Gamma_{\rm CR}(n)\, W/ v_A$.
Then, since all terms on the RHS of Eq.~\eqref{eq:Weq} are proportional to $v_A$, the stationary solution is independent of $v_A$, and by redefining $\tau = v_{A} k\, t$, we can write Eq.~\eqref{eq:Weq} as: 
\begin{equation}
\label{eq:Weq_2}
    \frac{\partial W}{\partial \tau}
    = c_k \frac{1}{k} \frac{\partial }{\partial k}\left[ k^{\frac{7}{2}} \sqrt{W} \frac{\partial W}{\partial k}\right]
    - \frac{1}{k}\frac{1}{r^2}\frac{\partial }{\partial r}(r^2 W) 
    + \frac{1}{k}Q^{(W)}
    .
\end{equation}
The replacement $\tau = v_{A} k\, t$ and its dependence on $k$ might be unexpected at first glance.
We note that this replacement does not allow to easily reconstruct the solutions for finite times $t$. But in the steady state solution ($t\rightarrow\infty$ and $\tau\rightarrow\infty$), where the LHS of the equation disappears, the solution for $W$ is identical. We verified this explicitly for a few examples.
The replacement allows faster convergence at small $k$.

The full discretized equation is given by (where each line corresponds to each term in Eq.~\eqref{eq:Weq_2}) : 
\begin{align}
    \ia{W}{t+1}{i}{j} &= \ia{W}{t}{i}{j}  \\
    & + \frac{c_k \Delta \tau }{k_j^2 (\Delta l)^2 } 
  \left[ 
     \RB{ \io{k}{j+\half} }^{\frac{5}{2}} 
     \sqrt{\left(\frac{\ia{W}{t}{i}{j}+\ia{W}{t}{i}{j+1}}{2} \right)}
     \RB{ \ia{W}{t}{i}{j+1}-\ia{W}{t}{i}{j} } 
     -\RB{ \io{k}{j-\half} }^{\frac{5}{2}} 
     \sqrt{\left(\frac{\ia{W}{t}{i}{j}+\ia{W}{t}{i}{j-1}}{2} \right)} 
     \RB{ \ia{W}{t}{i}{j}-\ia{W}{t}{i}{j-1} } 
  \right] \\ 
  & - \frac{\Delta \tau}{k_j} \SQB{ \frac{2}{r_i}  \ia{W}{t}{i}{j}
              + \frac{1}{\Delta{r_i}} \RB{ \ia{W}{t}{i}{j} - \ia{W}{t}{i-1}{j} } } \\
  & + \frac{\Delta \tau}{k_j} \ia{\RB{ Q^{(W)} }}{t}{i}{j}. 
\end{align}
where we have $W(\tau_t, r_i, l_j) = \ia{W}{t}{i}{j}$, and used a linear grid in space $r ( r_i) = \io{r}{i}$ (the same as for the density equation \eqref{eq:nediscr}) and a log grid in the wave number $k(l_j) = \io{k}{j}=\exp (l_j)$. 
We note that the wave number is related to energy since $E/{\rm GeV}=9.2\times 10^5\,(B_0/\mu{\rm G})({\rm pc}^{-1}/k)$. However, we solve Eq.~\eqref{eq:Weq} extending the gird in energy to avoid impacts from the boundary conditions (see below) imposed on $W$.

We defined $k_{j\pm\half} = \exp[ (l_j + l_{j\pm1} )/2] $.

Finally, we impose the following boundary conditions:
\begin{align}
  \partial_k W \RB{ r, k=k_{\rm min}, t} &= 0 &\longrightarrow&
  & \ia{W}{t}{i}{0} &= \ia{W}{t}{i}{1} \label{eq:W_BC_1} \\
  W \RB{ r=L, k, t} &= 0 &\longrightarrow&
  & \ia{W}{t}{n_r-1}{j} &= 0   \label{eq:W_BC_2} \\
  W \RB{ r, k=k_{\rm max}, t} &= 0 &\longrightarrow&
  & \ia{W}{t}{i}{n_k} &= 0   \label{eq:W_BC_3} \\
  \partial_r W \RB{ r=0, k, t} &= 0  &\longrightarrow&   
  & \ia{W}{t}{i}{-1} &= \ia{W}{t}{i}{0} && \label{eq:W_BC_4} 
\end{align}

We apply the Neumann boundary condition Eq.~\eqref{eq:W_BC_1} because the asymptotic solution for $k_{\rm min}\rightarrow 0$ in the diffusion-only scenario is a constant value $W$. However, this asymptotic behavior is broken by the advection term as can be seen for example in Fig.~\ref{fig:W_example}. The solution goes quickly to zero as $k$ becomes smaller than the resonant value corresponding to $E=m_{\rm DM}$. Since we set $k_{\rm min}$ much smaller than this value the boundary condition becomes irrelevant. 
Also for the boundary condition at $k_{\rm max}$ in Eq.~\eqref{eq:W_BC_3} we could have imposed the asymptotic behavior of the diffusion-only equation. On the other hand, we explicitly checked that this boundary condition affects the solution only down to $k_{\rm max}/\mathcal{O}(1)$, and we chose $k_{\rm max}$ corresponding to an energy much smaller than the ones used to derive the DM bounds. Thus, also the boundary condition at $k_{\rm max}$ becomes irrelevant and do not impact our results.

For the results shown in this paper, we set $E_{\rm min}=0.1$ GeV, $E_{\rm max}=m_{\rm DM}$, $k_{\rm min}=100\,{\rm pc}^{-1}$ and $k_{\rm max}=10^8\,{\rm pc}^{-1}$.

\section{Radio observations and data reduction}
\label{sec:obs}
\begin{figure}
\centering
   \includegraphics[width=1.0\textwidth]{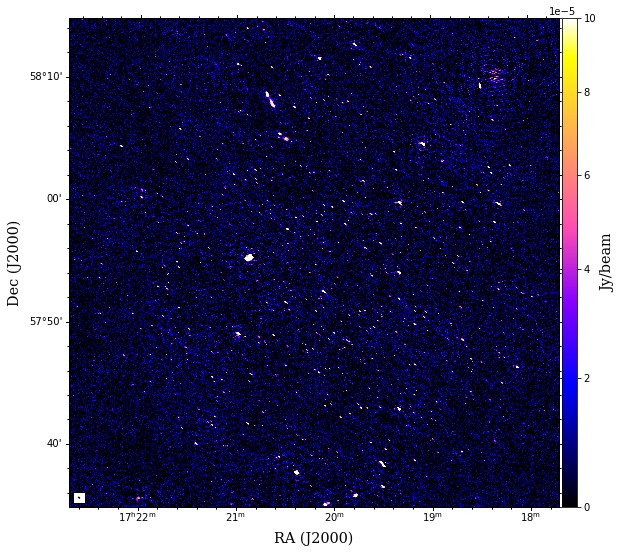}
    \caption{Image of the field centred on the Draco dSph galaxy with the uGMRT at 650 MHz. The angular resolution is $8'' \times 3''$ and the rms noise is 10~$\mu$Jy~beam$^{-1}$.}
\label{fig:draco}
 \end{figure}

Observations of the Draco dSph galaxy were carried out with the uGMRT at band~4. Details are reported in Table~\ref{tab:table_obs}. Observations were preceded by a 20~minute observation of the bandpass calibrator 3C\,286, assumed to be 21.4~Jy at 550~MHz, with a spectral index $\alpha = -0.85$ \citep{perley17}.
Data reduction was carried out with the Source Peeling and Atmospheric Modeling \citep[SPAM,][]{intema2009} pipeline. Given the wide (200~MHz) band, observations were split in four sub-bands of 50~MHz each and calibrated separately. The pipeline carries out the excision of radio frequency interference, followed by initial calibration and imaging of each sub-band individually. Images were further deconvolved individually in order to obtain a sky model for selfcalibration. Calibration solutions were specifically determined for bright, offending sources that were, therefore, accurately subtracted.
The calibrated sub-bands were eventually jointly imaged with \texttt{WSClean v3.1} \citep{offringa2014}, using a {\it robust~=~0} weighting scheme, yielding to a $8'' \times 3''$ synthesized beam. The deconvolved image is shown in Fig.~\ref{fig:draco}, with a 10~$\mu$Jy~beam$^{-1}$ rms noise.

The high resolution image was used to model all the compact sources brighter than 50~$\mu$Jy~beam$^{-1}$ which were subtracted from the visibilities. Residual visibilities were then imaged at lower resolution $32'' \times 15''$ and few more sources were identified and subtracted to obtain a final 32~$\mu$Jy~beam$^{-1}$ rms noise for the image that has been used to derive the bounds shown in the main text.

\begin{table*}[h]
\caption{Observation details.}
\centering
\begin{tabular}{@{}cccccc}
\hline
\hline
Target  & RA$_{\bf J2000}$  &   DEC$_{\bf J2000}$    & Frequency range    & Channel width 	& Integration time \\
    &   &   &   (MHz)   &   (kHz)   &   (hours) \\
\hline
Draco   &   $17^{\rm h} 20^{\rm m} 12^{\rm }.401$ &   $+57^{\circ} 20' 55''$ &  $550-750$    &   97.656   & 3.1 \\
\hline
\end{tabular}
\label{tab:table_obs}
\end{table*}

\section{Jeans analysis of the Draco dSph galaxy}
\label{sec:jeans}

As mentioned on Section \ref{sec:dSph}, in order to constrain the spatial distribution of the DM density, we performed a Jeans analysis using Draco's stellar kinematics. We used the line-of-sight (LOS) velocities sample from Ref.~\cite{2015MNRAS.448.2717W}. In addition to LOS velocities, this dataset contains information on effective temperatures, metallicities and surface gravities of stars, which has been used to select the final sample of LOS velocities, following the same procedure as in Ref.~\cite{Read:2018pft}. The LOS projected spherical radial velocity is \cite{10.1093/mnras/190.4.873,10.1093/mnras/200.2.361}
\begin{equation}
    \sigma_{\rm{LOS}}^{2}(R) = \frac{2}{\Sigma(R)}\int_{R}^{\infty}\left(1 - \beta \frac{R^2}{r^2} \right) \nu \sigma_{r}^{2} \frac{r dr}{\sqrt{r^2 - R^2}},
    \label{eq:sigma_los}
\end{equation}
where $\Sigma(R)$ is the projected luminous surface density, $\beta$ is the orbital velocity anisotropy, $\nu(r)$ is the stellar density and $\sigma_r(r)$ is the radial velocity distribution which can be obtained solving the spherical Jeans equation
\begin{equation}
    \frac{1}{\nu(r)}\frac{\partial}{\partial r} (\nu(r) \sigma_r^2) + \frac{2 \beta(r) \sigma_r^2}{r} = -\frac{G M(<r)}{r^2},
    \label{eq:jeans}
\end{equation}
where $M(<r)$ the mass content up to a radius $r$. 

Besides the kinematic data of the stars, a photometric sample is needed in order to estimate the stellar density distribution in Draco. It is taken from Ref.~\cite{2020ApJS..251....7F}. 
Then we model the stellar surface brightness density as a sum of three (projected) Plummer spheres \cite{10.1093/mnras/200.2.361}

\begin{figure}[b!]
\centering
\begin{minipage}{0.5\textwidth}
  \centering
  \includegraphics[width=1\linewidth]{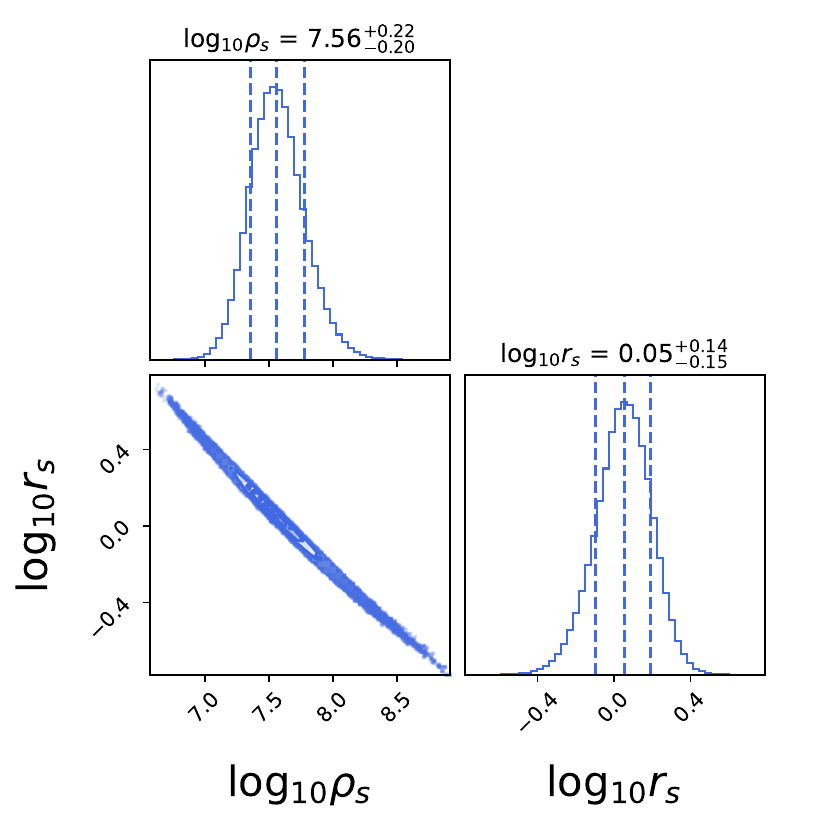}
\end{minipage}%
\begin{minipage}{0.5\textwidth}
  \centering
  \includegraphics[width=1\linewidth]{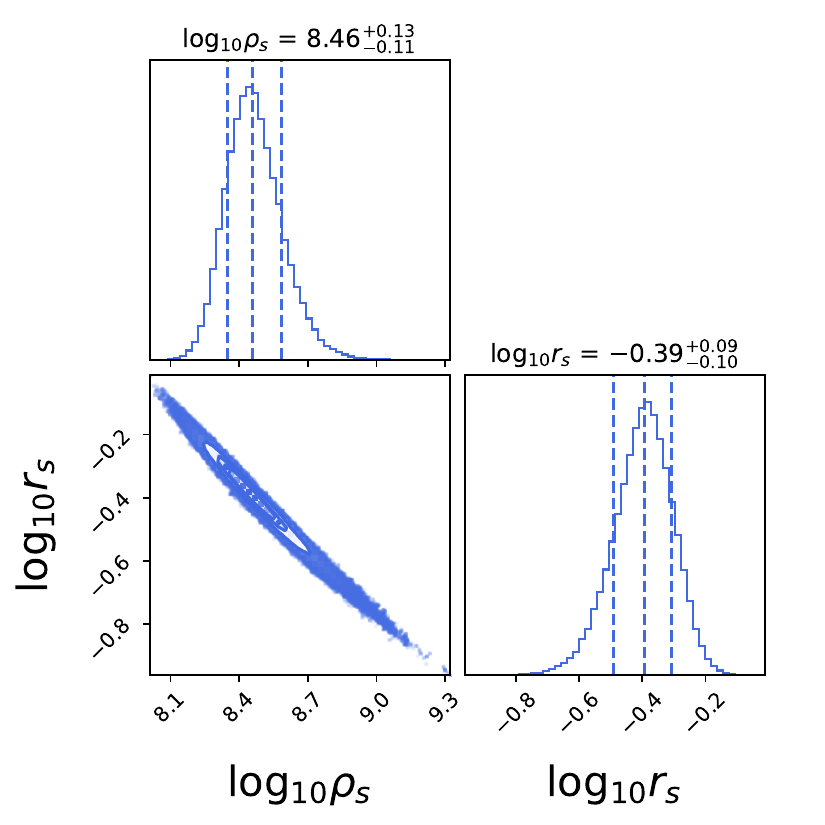}
\end{minipage}
\caption{Posterior distributions of the DM parameters $\log_{10}(\rho_s/(M_\odot/\rm{kcp}^3))$ and $\log_{10}(r_s/\rm{kpc})$ from equation \eqref{eq:dark_matter}. Left panel shows the results of the MCMC for the NFW profile. Right panel shows the results in the Burkert case.  }
\label{fig:triangle}
\end{figure}

\begin{equation}
    \nu(r)= \sum_{j=1}^{3}\frac{3 M_j}{4\pi a_j^3} \left( 1 + \frac{r^2}{a_j^2} \right)^{-5/2}
\;\;\;,\;\;\;
    \Sigma(R)= \sum_{j=1}^{3} \frac{M_j}{\pi a_j^2} \left( 1 + \frac{R^2}{a_j^2} \right)^{-2},
\label{eq:stellar_profile}
\end{equation}
with $M_j$ and $a_j$ being parameters that we include in the likelihood and estimate from the data. The contribution to the enclosed mass in Eq.~\eqref{eq:jeans} is computed through the integration of the stellar density, whilst the DM part is modeled either through a NFW \cite{navarro_1996} or a Burkert \cite{Burkert_1995} profile,
\begin{equation}
    \rho_{\rm{NFW}}(r) = \frac{\rho_s}{\left( \frac{r}{r_s}\right)\left( 1 + \frac{r}{r_s} \right)^2}\;\;\;,\;\;\; \rho_{\rm{Burkert}}(r)= \frac{\rho_s}{\left( 1+  \frac{r}{r_s} \right) \left( 1 + \left(\frac{r}{r_s}\right)^2\right)}. \label{eq:dark_matter}  
\end{equation}

In addition to the kinematic and photometric data, we also make use of higher velocity moments called Virial Shape Parameters (VSPs) which in principle can help alleviate the mass-anisotropy degeneracy \cite{2017MNRAS.471.4541R}, and are given by:
\begin{subequations}
    \begin{equation}
        v_{s1}=\frac{2}{5}\int_0^\infty G M \nu(r) \left( 5- 2\beta(r)\right)\sigma_r^2 rdr = \int_0^{\infty} \Sigma(R)\left \langle  v_{\rm{LOS}}^4\right\rangle R dR,
    \end{equation}
    \begin{equation}
   v_{s2}=  \frac{4}{35}\int_0^{\infty} G M \nu(r) \left( 7 - 6 \beta(r) \right)\sigma_r^2r^3dr = \int_0^{\infty} \Sigma(r) \left\langle v_{\rm{LOS}}^4 \right\rangle R^3 dR,
    \end{equation}
\end{subequations}
being the RHS of the equation obtained trough the kinematic data and the LHS the prediction of the model. With all these ingredients we perform an MCMC analysis using the python package {\sc emcee} \cite{2013PASP..125..306F} implemented through the non-parametric code {\sc Gravsphere} \cite{2017MNRAS.471.4541R} and the {\sc python} wrapper {\sc pyGravSphere} \cite{2020MNRAS.498..144G}. For more details on the method, see Ref.~\cite{Reynoso-Cordova:2022ojo}. Our results for the posterior distributions of the DM parameters $\log_{10}(\rho_s/[M_{\odot} / \mathrm{kpc}^3])$ and $\log_{10}(r_s/[\mathrm{kpc}])$ are shown in Fig.~\ref{fig:triangle}.

\begin{figure}[t!]
\vspace{-25mm}
   \includegraphics[width=0.45\textwidth,trim={0.4cm 0.79cm 3.0cm 3.5cm},clip]{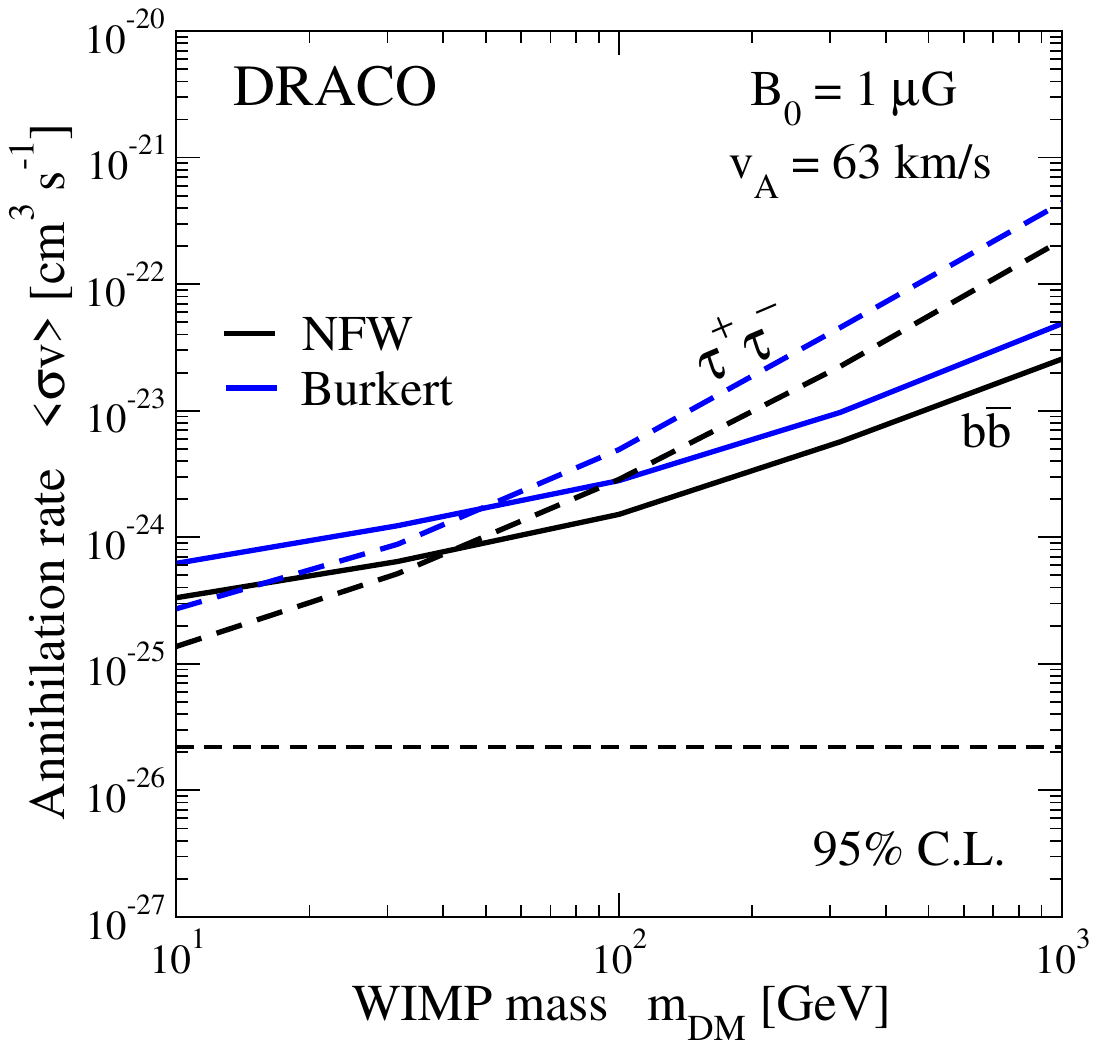}
    \caption{Similar to Fig.~\ref{fig:bounds} but comparing the bounds derived with the NFW and Burkert profiles. }
\label{fig:boundsBUR}
 \end{figure}

In Fig.~\ref{fig:boundsBUR}, we show a comparison between the bounds derived in Fig.~\ref{fig:bounds}, adopting an NFW profile, with the ones obtained for a Burkert profile. The difference between the two cases is limited. This can be understood from: a) the J-factor over $30'$ of the two models is very similar; b) the size of the Burkert core is significantly smaller than the diffusive region (and thus it has a limited impact in reducing the turbulence generation); c) the spatial diffusion is partially washing out the central cusp of the NFW profile.  

\section{Estimate of the diffusion coefficient with analytical approximations}
\label{sec:est}

In this Appendix we aim at deriving a simple (and clearly approximated) expression for the diffusion coefficient induced by the self-generation mechanism described in this work.

By neglecting advection and taking the stationary limit of Eq.~\eqref{eq:Weq}, one obtains
\begin{equation}
\label{eq:Weqsimple}
   \frac{\partial }{\partial k}\left[D_{kk}(W)\frac{\partial W}{\partial k}\right]
    =- \Gamma_{\rm CR}(n) W
     \; .
\end{equation}
Before computing the solution, notice that the l.h.s. is a damping term $\Gamma_d\,W$ with $\Gamma_d\sim D_{kk}/k^2$. For typical values of the parameters of the models described in the main text, the associated timescale at 1 GeV is $1/\Gamma_d\sim 10^5$ yr. 

To proceed towards the solution of Eq.~\ref{eq:Weqsimple}, let us now consider spectral and spatial features of the equilibrium distribution of the $e^+-e^-$ to be separable and assume a power-law for the energy spectrum, namely:
\begin{equation}
\label{eq:nesimple}
    n_e=10^{-12}\,\rm{GeV^{-1}\,cm^{-3}}\,\tilde{N}_0\,\left(\frac{E}{\rm GeV}\right)^{-3+\delta}\,\tilde{f}(r/r_*)\;,
\end{equation}
where $\tilde{N}_0$ is just a normalization constant, $\delta$ gives the spectral index of the energy distribution of $n_e$ and $\tilde{f}$ describes its spatial profile, that we will write in terms of a typical scale $r_*$ (which could be, e.g., the half-light radius or the NFW scale radius), where it is normalized $\tilde{f}(r=r_*)=1$.
Putting together Eqs.~\eqref{eq:Weqsimple} and \eqref{eq:nesimple}, taking the term $p^4\,f$ at the resonance, and after some simple numerical evaluations, one gets
\begin{equation}
\label{eq:Weqsimple2}
  - \frac{\partial }{\partial k}\left[k^3 (k\,W)^\gamma\frac{\partial W}{\partial k}\right]=3.3\times 10^{-8}\,\tilde{N}_0\, \left[\frac{(10^{-6}\,\rm{pc})^{-1}}{k}\right]^\delta\left(\frac{B_0}{\mu\rm{G}}\right)^{-3+\delta}\left|\frac{\partial\tilde{f}}{\partial r/(100\,\rm{pc})}\right|
        \; .
\end{equation}
By describing the turbulence with $W=W_0\,k^\beta$, and plugging it into Eq.~\eqref{eq:Weqsimple2},
it follows $\beta=-1-\delta/(\gamma+1)$ and the solution reads:
\begin{equation}
\label{eq:Weqsimple4}
  W=W_0\,k^{-1-\delta/(\gamma+1)}\;\;\;\rm{with}\;\;\;W_0=\left(\frac{3.3\times 10^{-8}}{(1-\delta)\,[1+\delta/(\gamma+1)]}\,\frac{\tilde{N}_0}{(10^{-6}\,\rm{pc})^\delta}\left(\frac{B_0}{\mu\rm{G}}\right)^{-3+\delta}\left|\frac{\partial\tilde{f}}{\partial r/(100\,\rm{pc})}\right|\right)^{1/(1+\gamma)}    
     \; .
\end{equation}
Notice that, as already discussed in the main text, the turbulence scales inversely with the strength of the coherent magnetic field, and directly with the spatial derivative of the distribution function. Finally, from Eq.~\eqref{eq:dc}, namely
$D\simeq c\,r_L/[3\,k_{res}\,W(k_{res})]$, we arrive at the expression of the diffusion coefficient:
\begin{equation}
    \label{eq:Dsimple}
  D=3.3\times 10^{22}\rm{cm^2/s}\,\left(\frac{3.3\times 10^{-8}\,\tilde{N}_0}{(1-\delta)\,[1+\delta/(\gamma+1)]}\,\left|\frac{\partial\tilde{f}}{\partial r/(100\,\rm{pc})}\right|\right)^{-\frac{1}{1+\gamma}}    \,\left(\frac{E}{\rm GeV}\right)\,\left(\frac{B_0}{\mu\rm{G}}\right)^{\frac{3-\delta}{1+\gamma}-1}\,\left(\frac{k_{res}}{(10^{-6}\,\rm{pc})^{-1}}\right)^{\frac{\delta}{\gamma+1}}
     \; .
\end{equation}
Taking now a typical example, i.e., $\delta=0$, and $\gamma=1/2$ (Kolmogorov), the estimate of the diffusion coefficient becomes
\begin{equation}
        D=3\times 10^{27}\rm{cm^2/s}\,\left(\tilde{N}_0\,\left|\frac{\partial\tilde{f}}{\partial r/(100\,\rm{pc})}\right|\right)^{-2/3}    \,\left(\frac{E}{\rm GeV}\right)\,\left(\frac{B_0}{\mu\rm{G}}\right)\;.
\end{equation}

From the above expression, it looks clear that for magnetic fields at $\mu$G level and GeV particles, the effect of self-confinement can be relevant if the $e^+-e^-$ energy density is $\gtrsim 10^{-12}\,\rm{GeV/cm^3}=10^{-3}\,\rm{eV/cm^3}$ and with a significant spatial variation of the distribution.

In the above estimate, we neglected advection. This is justified only for $D\gg 10^{27}\rm{cm^2/s}$ (see main text), otherwise the full equation has to be solved.
In order to have an estimate of $q_{CR}$ needed to have a significant turbulence production and so of the DM annihilation rate needed, one can solve Eq.~\eqref{eq:qc} by again taking the stationary limit and neglecting advection and energy losses (since in the physical cases considered here the associated time-scales are longer than the diffusion time-scale), i.e., solving $D\,\partial n_e/\partial r=-\int_0^r dr'\,r'^2\,q_{\rm CR}(r')/r^2$. It can be easily verified that this approach leads to similar results as the ones presented in the main text with a more sophisticated numerical treatment.

\end{document}